\documentclass[10pt,journal,compsoc]{IEEEtran}
\usepackage[nocompress]{cite}
\usepackage{algorithmic}
\usepackage{array}
\usepackage{url}
\usepackage{amsthm, amsmath,amssymb,amsfonts}
\usepackage{graphicx}
\usepackage{textcomp}
\usepackage{xcolor}
\usepackage{xspace}
\usepackage{booktabs}
\usepackage{tabulary}
\usepackage{balance} 
\usepackage{threeparttable}
\usepackage{multirow}
\usepackage{subcaption,booktabs}
\usepackage{hyperref}
\usepackage[utf8x]{inputenc}
\usepackage{enumitem}
\usepackage{ulem}
\usepackage{soul}
\usepackage{wasysym}
\usepackage{fdsymbol}

\newcommand{\nmark}[1]{\underline{\textbf{#1}}}
\newcommand{\etal}{\textit{et al}. }
\newcommand{\model}{NLTrace\xspace}

\newcommand{\pretrain}{pre-train\xspace}
\newcommand{\pretraining}{pre-training\xspace}
\newcommand{\pretrained}{pre-trained\xspace}

\newcommand{\bert}{\textbf{B}idirectional \textbf{E}ncoder \textbf{R}epresentations  from \textbf{T}ransformers\xspace}
\newcommand{\clssebert}{SE-BERT-CLS\xspace}
\newcommand{\clstask}{Task-CLS\xspace}
\newcommand{\clsproj}{Proj-CLS\xspace}
\newcommand{\ranktask}{Task-RANK\xspace}
\newcommand{\rankproj}{Proj-RANK\xspace}
\newcommand{\rankglue}{GLUE-RANK\xspace}
\newcommand{\edge}[2]{\text{⟨}#1,~#2\text{⟩}}
\newcolumntype{C}[1]{>{\centering\let\newline\\\arraybackslash}p{#1}} 

\newcommand{\PRE}{PRE}
\newcommand{\PAD}{PDA}
\newcommand{\ADJ}{ADJ}
\newcommand{\modelExplain}{\small { \PRE=Pretraining, \PAD=Project Adaptation, and \ADJ = Adjacent task}}

\newcommand*{\mybox}[2]{\colorbox{#1!30}{\parbox{.98\linewidth}{#2}}}
\definecolor{lightgray}{rgb}{.7, .8, .8}
\theoremstyle{definition}
\newtheorem{definition}{Definition}[section]

\begin{document}

\title{Enhancing Automated Software Traceability by Transfer Learning from Open-World Data}

%
%
%
%

\author{Jinfeng Lin, Amrit Poudel, Wenhao Yu, Qingkai Zeng, Meng Jiang, Jane Cleland-Huang }

\IEEEtitleabstractindextext{%
\begin{abstract}
Software requirements traceability is a critical component of the software engineering process, enabling activities such as requirements validation, compliance verification, and safety assurance. However, the cost and effort of manually creating a complete set of trace links across natural language artifacts such as requirements, design, and test-cases can be prohibitively expensive. Researchers have therefore proposed automated link-generation solutions primarily based on information-retrieval (IR) techniques; however, these solutions have failed to deliver the accuracy needed for full adoption in industrial projects. Improvements can be achieved using deep-learning traceability models; however, their efficacy is impeded by the limited size and availability of project-level artifacts and links to serve as training data. In this paper, we address this problem by proposing and evaluating several deep-learning approaches for text-to-text traceability. Our method, named \model, explores three transfer learning strategies that use datasets mined from open world platforms. Through pretraining Language Models (LMs) and leveraging adjacent tracing tasks, we demonstrate that \model can significantly improve the performance of LM based trace models when training links are available. In such scenarios \model outperforms the best performing classical IR method with an 188\% improvement in F2 score and 94.01\% in Mean Average Precision (MAP). It also outperforms the general LM based trace model by 7\% and 23\% for F2 and MAP respectively. In addition, \model can adapt to low-resource tracing scenarios where other LM models can not. The knowledge learned from adjacent tasks enables \model to outperform VSM models by 28\% F2 on generation challenges when presented with a small number of training examples. 
\end{abstract}}

\maketitle

\IEEEdisplaynontitleabstractindextext

%
\IEEEpeerreviewmaketitle

\IEEEraisesectionheading{\section{Introduction}
\label{tse_sec:introduction}}
\IEEEPARstart{S}{oftware} and systems requirements traceability is defined as `the ability to describe and follow the life of a requirement in both a forwards and backwards direction (i.e., from its origins, through its development and specification, to its subsequent deployment and use, and through periods of ongoing refinement and iteration in any of these phases)' \cite{DBLP:conf/re/GotelF94,DBLP:conf/re/GotelF95}. Traceability  establishes associations between different levels of requirements and with other types of artifacts, such as design specifications, test cases, models, and process descriptions. It supports numerous activities such as requirements validation and verification, safety assurance, and impact analysis, and is prescribed by many regulatory standards as part of the certification process \cite{do178b,ecss-e-40, IEC_60880_2013,conf/refsq/BouillonMP13, conf/re/MaederGP09, conf/refsq/RempelMKP13, conf/re/RempelMK13, journals/software/MaderJZC13}.

Given the high cost and effort required to manually create and maintain trace links during the software development process, researchers have proposed various solutions for generating links automatically  \cite{Gotel:2012fk,Hayes:AdvLinkGen,DBLP:books/daglib/p/LuciaMOP12}. Classical information retrieval (IR) solutions, such as the Vector Space Model (VSM) \cite{DBLP:journals/tse/HayesDS06}, Latent Dirichlet Analysis (LDA) \cite{DBLP:conf/re/DekhtyarHSHD07, DBLP:conf/icse/AsuncionAT10}, and Latent Semantic Indexing (LSI) \cite{DeLucia:ArtefManag} have been explored in-depth over the past decade, but have met a glass-ceiling in terms of achievable accuracy, with basic machine learning (ML) approaches  \cite{DBLP:conf/icse/Cleland-HuangCGE10,DBLP:conf/icsm/MillsEH18,DBLP:conf/icse/MillsH17,DBLP:conf/seke/SpanoudakisGZ03} suffering from similar fates. The primary impedance is caused by their lack of semantic analysis and of the textual artifacts.

Recent advances in natural language processing introduce potentially more effective approaches for automatically generating accurate trace links. Guo \etal \cite{guo2017semantically} proposed Deep Learning (DL) Trace Models, which leveraged a Recurrent Neural Network (RNN), to generate trace links between requirements and design definitions. However, their approach required large amounts of previously created links for training purposes. While Guo \etal demonstrated the potential for DL techniques to outperform IR ones, they concluded that much larger training datasets were needed in order to achieve satisfactory degrees of accuracy. In general, NLP-based tracing solutions require huge amounts of training data in order to perform well; however, in practice, software projects often lack sufficiently large sets of artifacts, including trace links, to support training of DL models. 

More recently NLP researchers have proposed DL methods based on \pretrained Language Models (LMs). In general, these methods use a two-step framework that involves \pretraining an LM, followed by fine-tuning the model. In the first step, the LM conducts a self-supervised training task to learn general knowledge about natural language including its vocabulary and grammar. Then in the second step, the model is fine-tuned to perform a specific task using supervised training \cite{wikipedia_2021}. Such approaches can significantly reduce the need for labeled training data during fine-tuning because the \pretrained LM can effectively transfer its knowledge for use in downstream tracing tasks performed at the project level. For this reason, this two-step process has been used extensively to address various text-to-text NLP tasks such as Machine Translation, Question Answering, Natural Language Inference, and Recognition of Textual Entailment \cite{devlin2018bert, liu2019xqa}. However, little work has investigated the application of such an approach to software traceability.

The task of generating trace links between natural language artifacts represents a domain-specific text-to-text task in which the trace model is trained to understand the semantic relatedness between pairs of software artifacts. There are two main motivations for utilizing a LM in the domain of software traceability. First, as Allan \etal \cite{allan2003challenges} state, LMs learn word distributions from a massive corpus of text data which can be combined with additional information to accomplish diverse tasks. Trace models supported by \pretrained LMs, leverage the knowledge they acquire from the larger training corpus, to understand the semantics of NL sentences. As such, their level of comprehension is far better than that of classical machine learning models. In our recent work, we showed that the use of a \pretrained LM significantly improved the accuracy of generated trace links when used to trace feature descriptions to source code in three large OSS \cite{ICSE21-Jinfeng}. Our DL approach achieved accuracy of 75\% to 99\%, measured using Mean Average Precision (MAP), as compared to  50\% to 70\% (MAP) when using VSM. However, we  \pretrained an LM model using millions of documented python methods, previously mined by Husain \etal \cite{husain2019codesearchnet} from OSS systems, and also leveraged `code search' as an additional fine-tuning task. The resulting DL architecture was dependent upon the presence of source code and not designed to efficiently support text-to-text requirements traceability as addressed in this paper. This is particularly problematic as much of the traceability prescribed by standards for developing systems in safety-critical domains (e.g.,\cite{safety.do178c ,ecss-e-40, IEC_60880_2013} \cite{Comar2009, Farail2006}) uses text-to-text traceability to show, for example, that requirements are satisfied by design specifications, or that hazards have been fully mitigated by requirements.

In other work \cite{lin2022information} we explored bilingual traceability in OSS projects in which issues and code (including comments and commit messages) were written in more than one human language. We leveraged a bilingual LM to mitigate the semantic and language gap and found that the LM-based trace model only outperformed translation-enhanced IR models on larger projects for which training examples were sufficient to fine-tune the LM. These results suggest that {\bf current LM-based tracing solutions underperform on small projects} or in new projects experiencing the cold start problem where few or even no training examples are available \cite{DBLP:conf/msr/0004RCRHV16}. 

In this paper, we seek to achieve improvements in accuracy for text-to-text tracing tasks by leveraging knowledge acquired from large sized open world data. We start by utilizing various \pretrained LM models, and then explore different transfer learning strategies to fine-tune the initial LM model for use in a specific software project. We refer to our approach as \model. Further, we focus on three specific tracing tasks -- namely \underline{\smash{link completion}} in which links are generated to fill gaps in the set of existing trace links,  \underline{\smash{link expansion}} in which links are generated for new artifacts as they are added to the project, and  \underline{\smash{link generation}} in which trace links are generated from scratch without the benefit of any initial links as training data.

We experimented with three transfer learning strategies applied within two model architectures, and found that \model, using task-level transference, achieved the best performance. It outperformed vanilla BERT-based models by 7\% and 23\% with respect to the F2-Measure (i.e., the harmonic mean of precision and recall) when applied to link completion and expansion tasks, and also performed well on the link generation task, where we found that when even 10 links were provided as training examples (i.e., few-shot), \model outperformed the best IR model by 28\% with respect to the F2-Measure. 
Our work therefore makes the following contributions:
\begin{itemize}
    \item We provide formal definitions of three unique tracing tasks related to link completion, link expansion, and link generation. These tasks have been alluded to previous papers but without formal definitions or systematic evaluations.
    \item  We build a large OSS tracing dataset by mining the GitHub archive from 2015-2021 to provide an external data source for supporting transfer learning in DL tracing models. We extract issues, commits, pull requests and user comments from the REST API dump and mine their associations using heuristic rules. 
    \item We systematically investigate the use of three different transfer learning strategies to identify and compare effective techniques for increasing the accuracy of NL-NL trace link completion, expansion, and generation tasks.
    \item We release a \pretrained LM, targeted at supporting diverse natural language software engineering tasks, into the public domain (cf.~Open Science at end of the paper). In this paper, we utilized this LM to \pretrain a LM and enhance the accuracy of trace link generation in traditional software project environments for three different tracing tasks.
    \item We build a framework for searching, retrieving, and parsing domain-specific documents, which can be used to construct a project-specific corpus, and show how this corpus can be used to increase tracing accuracy within a targeted domain.
    
\end{itemize}

The remainder of this paper is laid out as follows. In Section~\ref{tse_sec:problem}, we provide formal definitions of the tracing challenges and then in Section~\ref{tse_sec:RQs} describe the research questions addressed in this study.  Section~\ref{tse_sec:method} describes the data collection, transfer strategies and model architectures used in this study, and  Sections~\ref{tse_sec:exp} and \ref{tse_sec:results} introduce the experiment setup and discuss the results.  Finally, in Sections ~\ref{tse_sec:threat} to ~\ref{tse_sec:conclusion} we discuss threats to validity, present related work, and draw conclusions.

\section{Problem Definition}
\label{tse_sec:problem}
Most existing studies that address traceability automation focus on the task of `trace link generation'; however, in practice there are three different tracing tasks associated with trace link generation, trace link completion, and trace link expansion  \cite{guo2017semantically}. Our work explores the effectiveness of transfer learning solutions upon all three of these tasks, and therefore we start by providing clear descriptions and formal definitions for each of them. 

\subsection{Trace Link Completion}
Trace Link Completion (TLC) refers to scenarios in which project stakeholders have already established trace links for a subset of the existing software artifacts. However, given the non-trivial cost of creating and maintaining trace links, these links are often incomplete \cite{journals/software/MaderJZC13,DBLP:conf/icse/Cleland-HuangGHMZ14}. Link completion tasks seek to automate the process of generating the missing links. We formally define the problem as follows:

\begin{definition}[{\bf TLC}]
Given a software engineering project $P= (S, T, L')$ constituted by source artifacts $S$, target artifacts $T$ and an incomplete set of trace links  $L'$, an automated trace model is used to generate the missing true links $ \Delta L = \{\edge{s_i}{t_i}\ \notin L' | s_i \in S, t_i \in T\}$ to produce a  completed set of links where project $P = (S,T, L' \cup \Delta L)$.
\end{definition}

\subsection{Trace Link Expansion} 
Trace Link eXpansion (TLX) refers to scenarios in which a complete set of trace links have been created by project stakeholders for all existing artifacts; however, new artifacts are introduced as the project evolves. TLX focuses on constructing links between existing and emerging artifacts. We formally define this problem as follows:

\begin{definition}[{\bf TLX}]
Given a software engineering project $P= (S, T, L)$ and emerging source artifacts $\Delta S$; the trace model is used to automatically create a new link set $ \Delta L = \{\edge{s_i}{t_i}\ | s_i \in \Delta S, t_i \in T\}$ and to update the project as $P' = (S\cup \Delta S, T, L \cup \Delta L)$.
\end{definition}

The TLX task is related to, but different from the Trace Link Evolution (TLE) task, described in our prior work \cite{DBLP:journals/ese/RahimiC18,DBLP:conf/icsm/RahimiGC16}. Whereas TLX focuses on generating new links in response to new artifacts, TLE evolves existing trace links in response to system modifications. Evaluating \model's support for TLE is outside the scope of this paper.

\subsection{Trace Link Generation}
Trace Link Generation (TLG) is used when no existing trace links exist -- either in a new project or a legacy project without existing links.  TLG generates trace links from  scratch, and is formally defined as follows:

\begin{definition}[{\bf TLG}]
Given a software engineering project without existing trace links $P= (S, T)$, the trace model is used to automatically create a link set $ L = \{\edge{s_i}{t_i}\ | s_i \in S, t_i \in T\}$ in order to produce a `link complete' project $P= (S, T, L)$
\end{definition}

A special case of TLG leverages sample links that are explicitly elicited from project stakeholders as training examples. Our study investigates whether a small number of example links, can improve the performance of the TLG task. For experimental purposes we explored the use of 10 training links, and therefore refer to this approach as 10 shot trace link generation and define it as follows:

\begin{definition}[TLG - 10 shots]
Given a software engineering project with a few trace links $P= (S, T, L')$ where $|L'| = 10 $, automatically create link set $ L = \{\edge{s_i}{t_i}\ | s_i \in S, t_i \in T\}$ to produce `link complete' project $P= (S, T, L' \cup L)$
\end{definition}

The work we present in this paper aims to deliver significant improvements in accuracy for each of these three tracing tasks (i.e., TLC, TLX, and TLG) when compared to existing IR and ML tracing solutions. 

\section{Research Questions}
\label{tse_sec:RQs}
The \bert (BERT) language model was initially proposed by Devlin \etal as a language model for supporting diverse NLP tasks \cite{devlin2018bert}. BERT supports the transfer learning theory that a model trained on {\it upper-stream tasks} can acquire knowledge to improve its performance on downstream tasks. This is applicable to the traceability challenge where  TLC, TLX, and TLG represent downstream tracing tasks.

A recent traceability study used special LMs, \pretrained using an intermingled corpus of text and code\cite{feng2020codebert}, to trace from NL to code written in various programming languages (i.e., NL-PL). The results showed that  the intermingled corpus enabled semantic comprehension of PL artifacts and delivered quite accurate tracing results -- at least for the TLC tasks on which it was evaluated.
However, targeted LMs are not currently available in the area of NL-NL tracing and to our knowledge no well-trained LMs have been \pretrained with a software engineering related corpus.  Further, it is unclear whether existing general purpose LMs, that have been extensively and successfully used to support other NLP tasks would perform well on the NL-NL tracing problem. One reason that this might not be the case is due to the highly technical vocabulary that often characterizes the domains of software intensive systems projects.
Our first RQ therefore addresses the following question in order to establish a baseline for exploring transfer learning techniques.
\vspace{3pt}

\noindent\textbf{RQ1: How well does {\it \model} perform without the benefit of domain-specific transfer learning, and does it outperform classical IR trace models and other previously described DL tracing models?}\vspace{2pt}

Transfer learning has been recognized as an effective approach for improving model performance \cite{zhang2021survey, pan2009survey}. The underlying notion is that training a model to perform a set of similar secondary tasks in addition to their primary task (i.e., generating trace links), enables the model to improve  the performance of its primary task. However, there is an underlying assumption that the secondary and primary tasks are sufficiently related, so as to allow the DL model to effectively transfer the knowledge that it learns from the secondary task(s) to improve its performance on the primary task. 

\begin{figure}[t]
    \centering
    \includegraphics[width=0.95\linewidth]{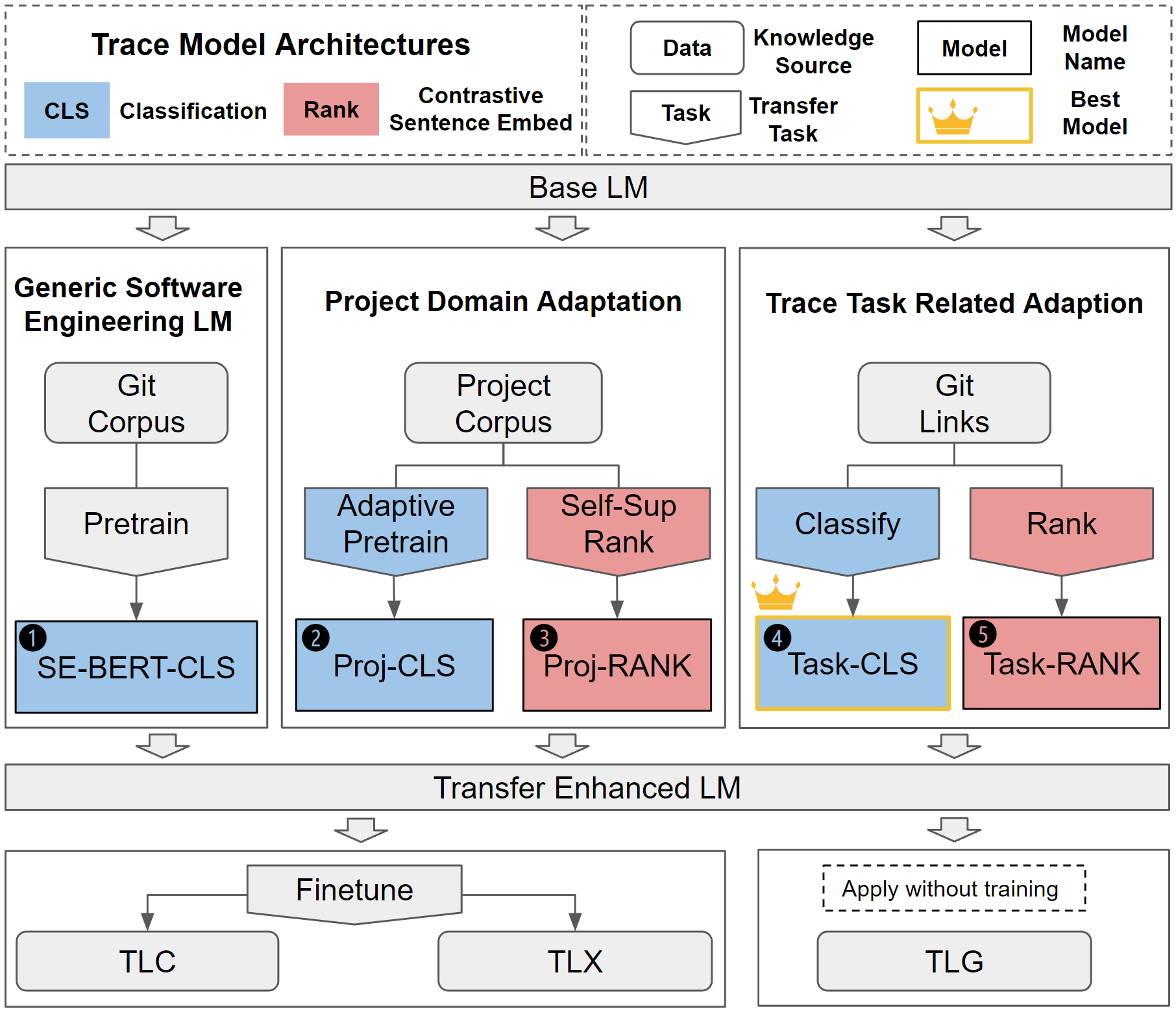}
    \caption{Our experiments evaluated \model with three transfer learning strategies including pretraining a  generic software engineering language model, adapting the LM to a specific project domain, and adaptation to the tracing task.  Three distinct sources of external knowledge were collected and applied to classification and sentence embedding based architectures, and ultimately five different \model variants (labeled 1-5) were evaluated. These models were used for three different traceability tasks, namely Trace Link Completion (TLC), Expansion (TLX), and Generation (TLG). }
 
    \label{tse_fig:models}
    \vspace{-5pt}
\end{figure}

As our goal is to adapt existing DL models to perform TLC, TLX, and TLG tracing tasks, we explore different combinations of model architectures and transfer learning tasks to determine which solutions can most effectively apply transfer learning techniques that utilize open world data sources to facilitate tracing in resource-limited software project environments. As depicted in   Fig.~\ref{tse_fig:models}, we explored three transfer strategies and evaluate them through a series of experiments. 

First, we built a LM targeted directly at the general Software Engineering domain by applying extensive pretraining strategies to a large SE related corpus. To this end, we collected more than 372GB of SE artifacts represented in plain text from millions of OSS projects on GitHub, and used the text data to pretrain a LM, which we named \emph{SE-BERT}. SE-BERT was able to learn the terminology used in requirements and design artifacts, more effectively than general LMs. 

Second, we adapted the \pretrained LM through additional pretraining based on domain specific text. This approach trained the LM model to better understand project-specific vocabulary by utilizing data from two sources that included (1) the artifacts and glossaries of the targeted project, and (2) documents retrieved using the Google Search Engine seeded with search queries extracted from terms and phrases in the project's artifacts.  

Finally, in our third approach, we explored the effectiveness of task-level knowledge transfer by formulating an adjacent tracing problem in which the model learned to  recreate a large set of hyperlinks that had previously been created by OSS project maintainers between GitHub pages \cite{autolink}. The updated model was then applied to our three tracing tasks. 

To explore the effectiveness of each of these techniques we formulated and addressed three additional research questions. In RQ2 we investigated the performance of the three proposed transfer-learning techniques used in conjunction with several different LM models. This part of our study focused on the two tracing tasks for which training data was available (i.e., TLC and TLX) and addressed RQ2, stated as:
\vspace{3pt}

\noindent\textbf{RQ2: Which, if any, of the three transfer learning strategies, produce LMs that outperform the original general-purpose LMs with respect to the TLC and TLX tracing tasks?}

To address the open challenge of the TLG problem, in which new trace links need to be generated without the benefit of existing training data we evaluated the benefit of providing a small number of training examples by asking:
\vspace{3pt}

\noindent\textbf{RQ3: Can LM models outperform classical IR methods on the TLG task when a small number of training examples are provided?}\\ \vspace{-4pt} 

Finally, we concluded our study by comparing results across these different methods. While an organization could use different tracing techniques for different tasks, there is significant benefit and reduced overhead if a single tracing model can be deployed for all tasks.  Therefore we addressed the final research question:
\vspace{3pt}

\noindent\textbf{RQ4: What is the overall best method for supporting all three NL tracing tasks?}

\section{Proposed Tracing Models}
\label{tse_sec:method}
\model leverages the well-known pretrain-then-finetune paradigm. It takes a general purpose BERT model as the starting point, conducts transfer learning tasks to improve the LM, and/or performs fine-tuning using the target project's data.  We evaluated the effectiveness of the three transfer learning strategies  within the context of two different model architectures. As shown in Fig.~ \ref{tse_fig:models}, we proposed and evaluated five  \model variants using the following techniques and components.

\subsection{Mining a Dataset as the LM Knowledge Source}
\label{tse_sec:data_pipeline}
\subsubsection{Github Dataset}
As of April 2022, GitHub hosted over 73 million developers and more than 200 million repositories \cite{wikipedia_2022}. In order to build software systems in a collaborative environment, GitHub developers produce a large amount of textual content including source code, bug reports, feature requests, technical discussions, and pull requests. This data serves as a potentially excellent resource for building LMs targeted at text-based traceability tasks in the Software Engineering domain.

Many well-maintained GitHub projects use Autolink \cite{autolink} to track and manage complex relationships across different artifacts. Autolink is a GitHub feature which automatically transforms long URLs into a standard abbreviated form. For example, a developer can conveniently add a string `\#1' in their commit message to refer to Issue No.1 within the same project repository, or could reference a specific commit in their issue discussion by adding a 7 digits SHA hash (e.g. `a5c3785'). Since these links have a uniform format, a regular expression can be used to mine them from the artifacts. Furthermore, the large quantity of hyperlinks between textual artifacts, makes them well suited for  supervising the training of an NL-NL tracing model, thereby potentially mitigating TLG's cold start problem. Throughout the remainder of this paper we refer to the mined text as the \textit{Git Corpus} and the mined autolinks as \textit{Git Links}. The Git Corpus was used to build a Generic Software Engineering LM for use with a classification model (cf.,~Model \#1, Fig.~\ref{tse_fig:models}), while the Git Links were used for task level adaptation techniques (cf.,~Models \#4 and \#5).

\begin{figure}[t]
    \centering
    \includegraphics[width=0.95\linewidth]{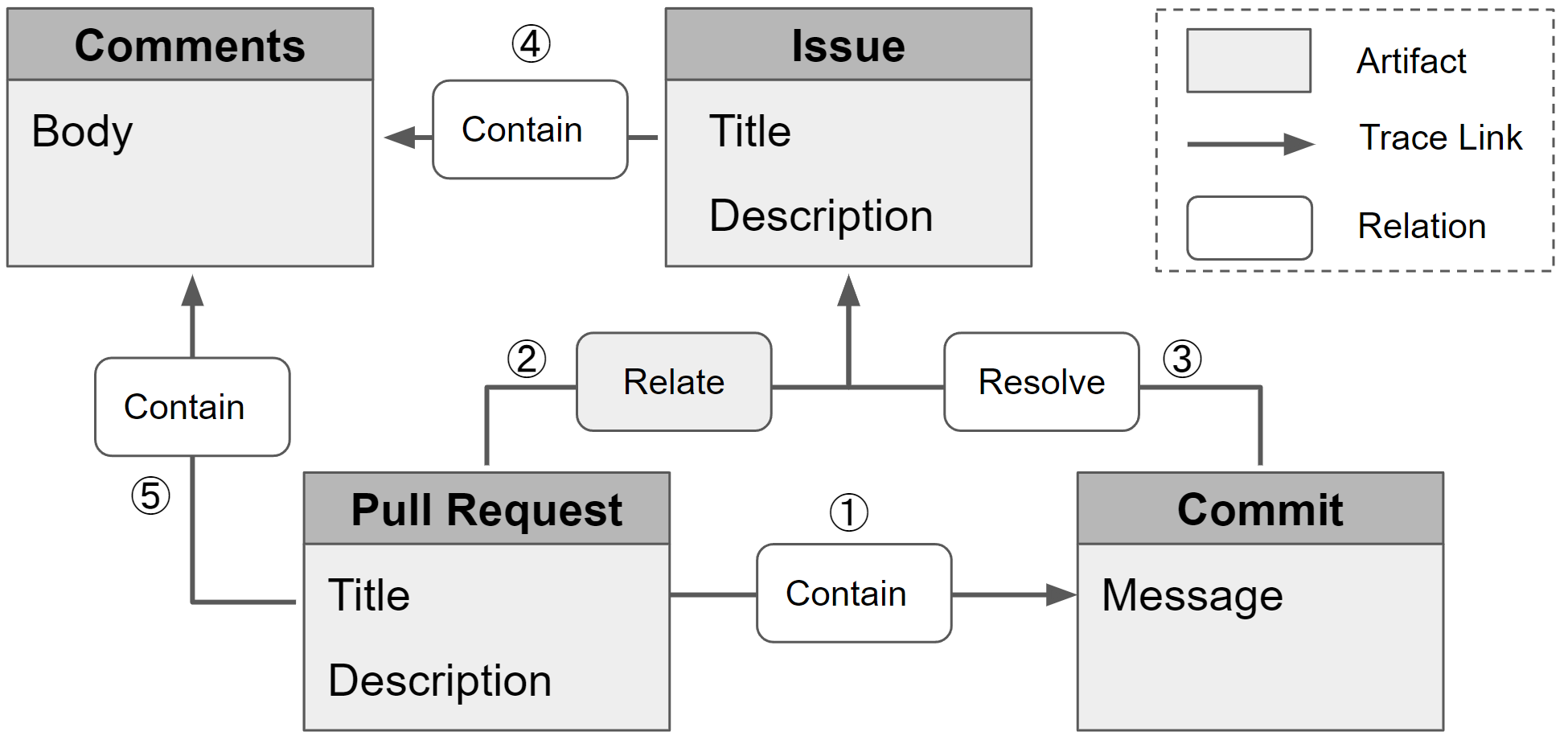}
    \caption{The TIM (Trace Information Model) used in conjunction with GitHub projects contains four types of artifacts and five types of links.}
    \label{tse_fig:git_data}
\end{figure}

For experimental purposes, we retrieved the Git Corpus and Git Links archived from 2016 to 2021, using the public API of the GH Archive project \cite{gh_archive} which returns all HTTP requests sent to the GitHub API during this period, in a standard JSON format \cite{github_docs}.  While data is available from 2011, we did not use it, because the format was not standardized until 2016.  The six years of downloaded data produced a 2.1TB zip file, which was used to reconstruct all repositories by parsing the HTTP requests ordered sequentially by date and time. We extracted four types of records from the requests including Comments, Issues, Pull Request and Commits, but found that Autolinks were primarily present in Issue-Commit and Issue-Pull requests. A Pull Request is usually associated with one or more Commits as its purpose is to deliver a patch that addresses a specific issue, while comments are hierarchically associated with Issues and Pull Requests. Their links can be obtained by parsing the JSON structure of the request payload. We abstracted the relationships among these four GitHub Artifacts depicted by the TIM (Traceability Information Model) shown in Fig.~\ref{tse_fig:git_data}. 

To effectively process such a large amount of data, we deployed a data pipeline on HTCondor, which is a distributed high throughput computation platform developed by Thain \etal \cite{thain2005distributed}. We harnessed a machine pool with 300 servers and distributed the workload evenly across these machines. We also processed the text to remove non-ascii tokens, code blocks and stack traces represented in Markdown format. We removed all artifacts with fewer than ten tokens after preprocessing, as they tended to be too short to provide meaningful content, following the cleaning process, and then verified that the endpoints of each link existed in the dataset. Any link with a missing endpoint was removed. Following this processing step, we obtained a Git dataset composed of the corpus and links with a size of 372GB.

\subsubsection{Domain-Specific Corpus Construction}
\label{tse_sec:google_searach_engine}
\begin{figure}[t]
    \centering
    \includegraphics[width=0.95\linewidth]{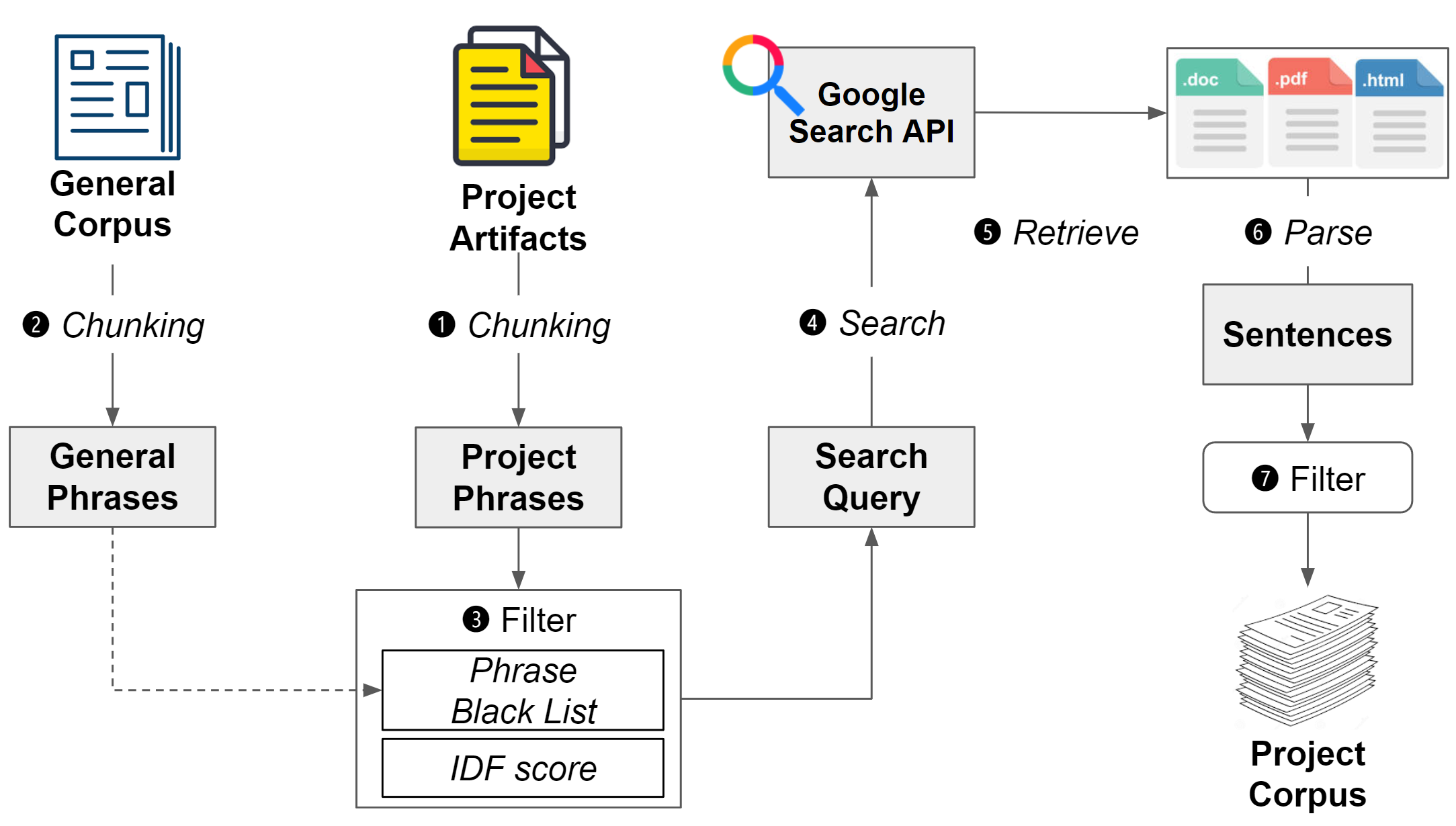}
    \caption{An automated data pipeline was used to extract concepts from target projects and to retrieve a  related corpus.}
    \label{tse_fig:gse_search}
\end{figure}

In addition to mining a corpus from GitHub, we also used the Google Search Engine to construct a domain-specific corpus for each of our target projects. Software artifacts usually contain technical terminology and jargon rarely found in vernacular language; therefore, we hypothesized that providing sentences that included those terms as training examples for LMs could potentially improve their performance on downstream tracing tasks. We therefore developed the pipeline depicted in Fig.~\ref{tse_fig:gse_search}, which first identifies domain-specific terms for a specific project, and then uses the Google Search Engine to retrieve contextualized examples of their use.  Starting with a project artifact (e.g., a requirement or design definition), we used NLTK \cite{bird2004nltk} to perform Noun Phrase Chunking in order to identify all noun phrases. These noun phrases included a mix of domain-specific phrases as well as more general concepts; however, building a corpus that includes general concepts could make its use in the transfer learning process less effective as the model is optimized for broad, and potentially irrelevant, content. We therefore generated a black list of general concepts, using the UMBC webBase corpus provided by Han \cite{UMBC_EBIQUITY_CORE_Semantic_Textual_Similarity_Systems}. This corpus contains 100 million web pages from more than 50,000 websites that were collected as part of the Stanford WebBase project in 2017, and has a compressed size of 13GB. We applied the same chunking methods to extract noun phrases from the UMBC corpus, ranked them by their frequency, and used the top 1\% of the list to create a black list composed of approximately 30k commonly used phrases. After filtering each of our domain-specific concept lists to remove terms in the black list, we calculated the importance of the remaining concepts by computing an IDF (Inverse Document Frequency) score within the project artifacts. Finally, we applied a manually defined threshold to remove additional concepts and produce a final list of domain-specific concepts for each project.

We then used the terms from each concept in turn to seed a query using the public API of the Google Search Engine. The returned URLs primarily linked to HTML pages, PDF files, and/or doc files, which we converted into plain text and tokenized into sentences, or long phrases such as those found in bulleted lists, using the NLTK sentence tokenizer. Finally, we discarded all sentences that did not have at least one token that overlapped with the query. The domain-specific corpus was used by Models \#2 and \#3 as shown in Fig. \ref{tse_fig:models}.

\subsection{Model Architectures}
Traceability tasks are typically formulated as either a classification problem or as a distributed representation learning problem. In the first case, binary classification models compute a confidence score which is used to predict whether a candidate trace link is a true link or not; while in the second case, a model projects the artifacts into a latent semantic space and then calculates similarity scores as distances between the source and target artifacts in this space. For purposes of our experiments, we selected a state-of-the-art architecture from each of these solution spaces, to serve as a comparative baseline for evaluating \model. 

\subsubsection{Classification Model}
We utilized a Single-BERT architecture from the TBERT framework proposed by Lin \etal \cite{ICSE21-Jinfeng} as our classification architecture. This model was selected as it was shown to perform well on the tracing problem, in comparison to alternative Twin and Siamese models. 
It first forms pairwise concatenations for source and target artifacts by adding a special separator token `[SEP]' between them, and then uses a single BERT model to encode the merged long sequence.

The BERT model use a self-attention mechanism to exploit the relevance between the vocabularies and to create a high dimension hidden-state matrix as output. The classification header, which is a 3-layer neural network takes this matrix as input and produces a score between 0 and 1 to indicate the relevance of the paired input. It then optimizes for standard Binary Cross Entropy Loss as the objective function \cite{loss_func} written as follows:
\begin{equation}
    L = - (ylog(p) + (1-y)log(1-p))
\end{equation}
where y is the label for a given training link and p is the predicted likelihood of it being a true link. We followed the setup of TBERT to create a balanced training dataset in which positive examples were over-sampled to achieve the same number as the negative examples. 

\subsubsection{Sentence Embedding Model}
\label{tse_sec:model_arch_cse}
Distributed representation learning approaches transform textual documents into vectors in latent space where the distance between pairs of vectors reflects their semantic relatedness. Classical IR methods such as VSM, LDA and LSI all use these types of vectors. For example, the source and target artifacts are vectorized based on TF (Term Frequency) and IDF scores, and then a function (e.g., Cosine Similarity) is applied to obtain a similarity score for the embedded artifacts. Candidate links are then typically ranked by their scores and filtered according to a threshold score in order to identify a set of candidate true links. 

We adopted the SimCSE framework \cite{gao2021simcse}, which is a BERT based sentence embedding model that uses contrastive learning tasks to calibrate the distance between artifacts. Contrastive learning is an unsupervised technique in which a model learns to differentiate between similar and dissimilar elements. It typically starts with an element (such as an image or a sentence), modifies it (e.g., via cropping, transforming, or replacing a word or phrase), and assumes that the modified version is similar to the original version, whereas elements that did not originate from the original element are dissimilar \cite{chen2020simple}. 
Although no previous work has applied SimCSE to the traceability tasks, it has been used in similar NLP problems and fits the traceability task well. We utilized SimCSE to perform both an externally supervised and a self-supervised task. 

For the externally supervised task, the model is trained to perform a set of small ranking tasks. For example, we applied it to the traceability task as follows. Given four source artifacts, four target artifacts, and four previously defined trace links between any pair of these source and target artifacts, SimCSE  learns to link each source and target artifact to produce a ranked list of the 16 potential pairs, such that the top 4 are marked as true links and the remaining 12 as non-links.  The model is then optimized to reduce the cosine distance between positive links and to increase the distance for negative examples. 

SimCSE incorporates the Cosine Similarity directly into the objective function as follows:
\begin{equation}
    \label{tse_eq:sime_cse_obj}
    L = - log \frac{e^{sim(h_i,h_i^{+})/\tau} }{
    \sum_{j=1}^{N} ( e^{sim(h_i,h_i^{+})/\tau} +  e^{sim(h_i,h_i^{-})/\tau} )
    }
\end{equation}
where $sim(h_i,h_i^{+})$ refers to the Cosine Similarity between $h_i$ and $h_i^+$, which are the hidden state vectors for artifacts in the true links; whilst $sim(h_i,h_i^{-})$ refers to the  Cosine Similarity between vectors of artifacts in negative links. 

For the self-supervised task, we leveraged SimCSE's ability to perform contrastive learning by creating multiple vector representations for a single sentence. Within the SimCSE neural network, the internal semantic representation, known as its hidden state, was applied to multiple dropout layers to randomly corrupt a small part of this representation. These dropout layers produce vectors that differ from each other by one or few digits. Pairs of vectors generated from the same original sentence were treated as links (i.e., positive examples), while all others were treated as non-links (i.e., negative examples). 

SimCSE is a multi-task model. For both the supervised and self-supervised approaches, it performs the link prediction and a Mask Language Modeling (MLM) task simultaneously. MLM objectives are weighted and appended to the ranking objective formulated in Eq.~\ref{tse_eq:sime_cse_obj}. For this reason, we did not create a SimCSE based counter-part for our model \#1 in Fig.~\ref{tse_fig:models}, because model \#5 already includes it as a sub-task.

\subsection{Transfer learning for Traceability}
\label{sec:transfer_strategies}
Many previous studies have been conducted to improve the ability of a LM to support downstream tasks. From these we identified the following three key strategies.

\subsubsection{Pretraining the LM for the SE Domain}
\label{sec:domainLM}
Our first transfer learning strategy focused on pretraining a {\bf Generic Software Engineering LM} to produce SE-BERT-CLS (Model 1 in Fig.~ \ref{tse_fig:models}).  It applied MLM  \cite{devlin2018bert} to an otherwise unorganized text corpus to empower multi-layer transformers to perform the tracing task. The pretraining procedure leveraged the model's attention mechanism to learn mutual relations between the hidden tokens and their surrounding context. 

Other researchers have applied MLM mechanisms to similar NLP problems. For example, in RoBERTa, Liu \etal \cite{liu2019roberta} optimized the pretraining process by introducing a dynamic MLM mechanism whilst leveraging an additional 16GB of text from BOOKCOR-PUS \cite{zhu2015aligning} and English Wikipedia in the pretraining process. They showed that RoBERTa outperformed BERT on the majority of general NLP tasks defined by the GLUE dataset \cite{wang2018glue}. In addition, targeted LMs have been created that adapt vanilla BERT to support specific domains and their associated tasks, For example,  Finbert \cite{araci2019finbert}, BioBERT \cite{lee2020biobert}, ClinicalBert \cite{huang2019clinicalbert} and SciBERT \cite{beltagy2019scibert} conduct continual pretraining on vanilla BERT to establish domain dedicated LMs for Finance, Biology, Medical Care and general Sciences respectively. Gururangan \etal \cite{gururangan2020don} referred to this type of technique as DAPT (Domain Adaptive Pretraining), in contrast to the TAPT (Task Adaptive Pretraining) which we describe in Sec~\ref{tse_sec:tapt}. This extensive body of work shows that  DAPT can benefit diverse domain-specific downstream tasks, and therefore potentially be useful for our tracing tasks.

However, DAPT requires a large sized domain-specific text corpus. For example, Gururangan \etal utilized around 40GB of text for each of four different experimental domains. While this might be feasible in the Software Engineering domain for large companies with a huge corpus of their own data, it is likely infeasible for most organizations to provide such quantities of data. For this reason, we use data mined from public data sources, as previously described in Section \ref{tse_sec:data_pipeline}, to pretrain SE-BERT.  We selected the `bert-base-uncased' LM \cite{bert-base-uncased} as our starting point, and applied dynamic MLM as the pretraining task for adapting the BERT model for tracing SE artifacts. We refer to this \pretrained model as SE-BERT.  We decided not to pretrain SE-BERT from scratch, based on observations made by the SciBERT team \cite{beltagy2019scibert}, who reported that training from scratch improved  performance by an average of only 0.7\%, but took a significant amount of time and computing resources.  SE-BERT is applied on Model \#1 in Fig. \ref{tse_fig:models}.

\subsubsection{Adapting the baseline LM to the Project Domain}
\label{tse_sec:tapt}
Our second transfer learning strategy, which we name {\bf Project Domain Adaptation}, focuses on project-level adaptation based on vocabulary and concepts of the project itself. This adopts  Task Adaptive Pretraining (TAPT), proposed  by
Gururangan \etal \cite{gururangan2020don} as an alternative of DAPT. Instead of building a large corpus for pretraining a domain-targeted LM, as in the case of \clssebert, they created a much smaller corpus targeted directly at the specific NLP task, which in our case includes the three tracing tasks of TLC, TLX, and TLG. In sufficiently large projects or organizations, the corpus could be assembled from internal documents and Wiki pages using product-specific terminology and jargon to describe products and processes  products; however, in other projects, external data sources need to be used.

As depicted in Fig.~\ref{tse_fig:models}, we created two different models named \clsproj (Model \#2) and \rankproj (Model \#3), based on classification and CSE architectures respectively. In both cases, we used the data pipeline described in Section~\ref{tse_sec:google_searach_engine} to automatically mine a project related corpus from open source datasets. 
For the \clsproj model, we used the MLM task to adapt the vanilla BERT with project dedicated vocabulary; while for \rankproj, we conducted a self-supervised ranking task as discussed in Sec.~\ref{tse_sec:model_arch_cse} to learn the languages in the projects.

\subsubsection{Learning From Adjacent Tasks}
\label{sec:tasklevel}
The final approach, which we refer to as {\bf Trace Task-Related Adaptation} adapts the LM by training it to perform similar (adjacent) tasks. As depicted in Fig.~\ref{tse_fig:models}, we applied adjacent task training within the context of the classification (Model \#4) and CSE (Model \#5) architectures. We previously showed that adjacent training tasks can be used effectively to generate trace links between NL and programming language (PL) artifacts \cite{lin2021traceability}. In that case we used code search as an adjacent task, and showed that transfer learning improved MAP scores by more than 20\%. We first built a BERT-based trace model to predict the relevance between python function doc strings and function specifications, and then fine-tuned it using links between Issue descriptions and the Code Change Set within Commit messages. Even though the format and content of artifacts in the code search were distinctly different from the tracing task, the model was able to effectively learn a general set of rules that improved the NL-PL tracing accuracy.

We therefore sought to identify effective adjacent tasks that could be applied to the NL-NL tracing task prior to fine-tuning the trace model. As shown in Fig.~\ref{tse_fig:git_data}, our git data contains five types of links between four types of git artifacts, of which the Issue-Commit links are most similar to trace links in SE projects. In traditional software and systems engineering projects, artifacts such as requirements and design specifications, are typically arranged into a hierarchical structure, in which high-level artifacts are refined into more detailed lower-level ones. This is similar to, but not the same as, the relationships between an Issue and its associated Commits. GitHub users typically create an Issue describing a problem or new feature request, and articulate their concrete implementations through one or more corresponding Commit messages. While the Pull Request-Commit links share similar characteristics, users sometimes simply copy the Commit message as a Pull Request description, making this type of link less valuable than Issue-Commits as an adjacent training task. Other types of associations, such as links from pull requests to issues, are closer to peer-to-peer relations, and therefore less representative of our targeted tracing tasks. Similarly we opted not to use links from comments to other artifacts due to the broad range of topics covered by the comments. Focusing only on the targeted types of links reduced the pretraining corpus to around 110GB. Our decision to use issue-commit links was supported by an informal experiment in which models trained with Issue-Pull links returned significantly fewer improvements than models based on Issue-Commit links. Given the OSS issue-commit links as a resource, we investigated various adjacent tracing tasks and their ability to support knowledge transfer into more traditional systems projects.

\section{Transfer-Learning Tracing Experiments}
\label{tse_sec:exp}
\subsection{Datasets}
We evaluated \model against four datasets as shown in Table~ \ref{tse_tab:dataset}. In selecting the datasets we established inclusion criteria that each dataset must include (i) traditional software requirement artifacts plus one additional NL artifact type (e.g., design specifications or regulations), and (ii) have an existing trace matrix containing at least 100 manually vetted links between the two artifact types. In addition (iii) we sought to select datasets from diverse domains. We included CM1 as an exception case as it represented a small project from a niche domain. Our criteria were defined to focus the research upon requirements traceability solutions in projects that follow a more traditional requirements-driven approach, as is common in safety-critical systems domains. This excluded the  use of OSS datasets that use informal feature requests in lieu of more traditional requirements, and therefore limited the number of datasets available to us. 
\begin{table}[tbh!]
	\centering
    \caption{Software Engineering projects in four domains. The projects were selected because they were non-trivially sized and provided manually created trace links that were used for validation purposes.}
    \begin{tabular}{llll}

    \toprule \hline
        Project  & Description & Source & Target\\ \hline
        PTC   & Subway signalling system & SRS & SDS \\ 
        CCHIT  & Electronic record system &Regulations& SRS\\ 
        Dronology& Multi-UAV flight coord. &SRS&SDS \\ 
        CM1& Scientific instrument &SRS&SDS \\ 
        \hline
        \bottomrule
    \end{tabular}
    \label{tse_tab:dataset}
\end{table}
\begin{table*}[t]
    \caption{ Train, development and test dataset splits for each project. Split-by-link and Split-by-artifacts were used for link completion and link expansion tasks respectively. For link generation, we utilized the Split-by-artifacts, but without use of links in the training set.}
    \centering
        \addtolength{\tabcolsep}{1pt}
        \begin{tabular}{cc|ccc|ccc|ccc|ccc} \toprule \hline 
        & & \multicolumn{3}{c|}{{\bf CCHIT}} & \multicolumn{3}{c|}{{\bf PTC}} & \multicolumn{3}{c|}{{\bf Drone}} & \multicolumn{3}{c}{{\bf CM1}} \\
        & & train & valid & test & train & valid & test & train & valid & test & train & valid & test \\ \hline
        \multirow{3}{*}{ Completion } & Source & 419 & 419 & 419 & 72 & 72 & 72 & 94 & 94 & 94 & 22 & 22 & 22 \\
        & Target & 1816 & 1816 & 1816 & 415 & 415 & 415 & 210 & 210 & 210 & 53 & 53 & 53 \\
        & True Links & 3241 & 405 & 406 & 470 & 58 & 59 & 167 & 21 & 22 & 22 & 10 & 13 \\ \hline
        \multirow{3}{*}{ Expansion } & Source & 335 & 41 & 43 & 57 & 7 & 8 & 75 & 9 & 10 & 7 & 7 & 8 \\
        & Target & 1816 & 1816 & 1816 & 415 & 415 & 415 & 210 & 210 & 210 & 53 & 53 & 53 \\
        & True Links & 3193 & 318 & 541 & 468 & 26 & 93 & 166 & 19 & 26 & 18 & 13 & 14 \\ \hline
        \multirow{3}{*}{ Generation } & Source & 335 & 41 & 43 & 57 & 7 & 8 & 75 & 9 & 10 & 7 & 7 & 8 \\
        & Target & 1816 & 1816 & 1816 & 415 & 415 & 415 & 210 & 210 & 210 & 53 & 53 & 53 \\
        & True Links & 0 & 318 & 541 & 0 & 26 & 93 & 0 & 19 & 26 & 0 & 13 & 14 \\ \hline
        \end{tabular}

    \label{tse_tab:data_detail}
\end{table*}

Our first dataset was a Positive Train Control (PTC) system, which supports communication and signaling for a large underground railway system. The dataset, including requirements, design specifications, and an associated trace matrix, was provided by our industrial collaborators under a non-disclosure agreement.  
The Dronology dataset \cite{cleland2018dronology} was developed at the University of Notre Dame for coordinating emergency response missions of multiple small unmanned aerial vehicles (UAVs). It includes over 10,000 LoC, and was developed by a mix of professional developers, post-docs, and both graduate and supervised undergraduate students. It has been used by over 20 external research teams to support research in areas such as product lines, security, and traceability \cite{rahimi2018evolving, krismayer2019constraint, krismayer2019supporting}. For purposes of this paper, we used a subset of the Dronology dataset including NL requirements, design definitions, and associated trace links. 
In both PTC and Dronology projects, the trace links were constructed by the original developers to support activities such as requirements validation and impact analysis. 
The CCHIT dataset was initially derived from two industrial sources and is available via COEST.org. It includes two sets of requirements. The first set was provided by the Certification Commission for Health Information Technology (CCHIT) for certifying electronic health records (EHRs) and the networks they use. 
The second set of requirements was provided by the Veteran Administration's Electronic Health Record system (WorldVista). Trace links in the CCHIT dataset are primarily used to support compliance analysis and were created by researchers for use in a prior publication \cite{DBLP:conf/icse/Cleland-HuangCGE10}. The CCHIT dataset is also available at COEST.org. 
The CM1 dataset is provided by NASA.  It is an extract from the artifacts of an interstellar telescope and includes high-level and low-level design requirements. The links between the artifacts were manually created by experts in NASA.

\subsection{Experiment Setup}
As discussed in Sec.~\ref{tse_sec:problem}, we focused on the three tracing tasks of TLC, TLX and TLG. To validate results from our experiments, we compared the links generated by our \model variants against the manually created trace links (aka the `answer set'), provided with each dataset. We split the datasets in distinct ways that were appropriate for each tracing task in order to create a training ({\it train}), validation ({\it valid}), and test ({\it test}) dataset.  

For the TLC task, we followed the `split-by-link' strategy adopted by Guo \etal \cite{guo2017semantically}. We performed a pairwise mapping between each source and target artifact to create the complete set of pairs, and then tagged them as `true' or `false' links according to how they were marked in the answer set. We then randomly split the candidate links into ten-folds, assigning eight folds for training, one for validation, and one for test. For this link completion task, all existing source and target artifacts were visible for each phase of training, validation, and testing. As CM1 has fewer links, we split its data into 2/1/1 train, validate, and test folds to ensure sufficient test links. By applying \model to predict the links in the test dataset, we simulated the trace link completion scenario in practice. 

For the TLX link expansion task, we adopted a  `split-by-artifact' method in which the source artifacts were randomly divided into ten-folds with 8/1/1 fold(s) assigned to train/dev/test sets respectively. Target artifacts were not divided, and all target artifacts were visible to each set of source artifacts.  This simulated the case in which a relatively complete set of target artifacts are available (e.g., requirements) whilst source artifacts (e.g., design specifications) are added over time. Within each split, we performed a pairwise mapping of the partial set of source artifacts to all target artifacts, and tagged positive and negative links according to the answer set. In this case, we simulated the expansion scenario in which 10\% of new source artifacts were added during software development and \model was used to predict the links between these new source artifacts and the existing target artifacts. For the smaller CM1 dataset, we assigned an equal number of source artifacts in each partition, leading to a test set with 14 links.

\begin{figure}[t]
    \centering
    \includegraphics[width=0.95\linewidth]{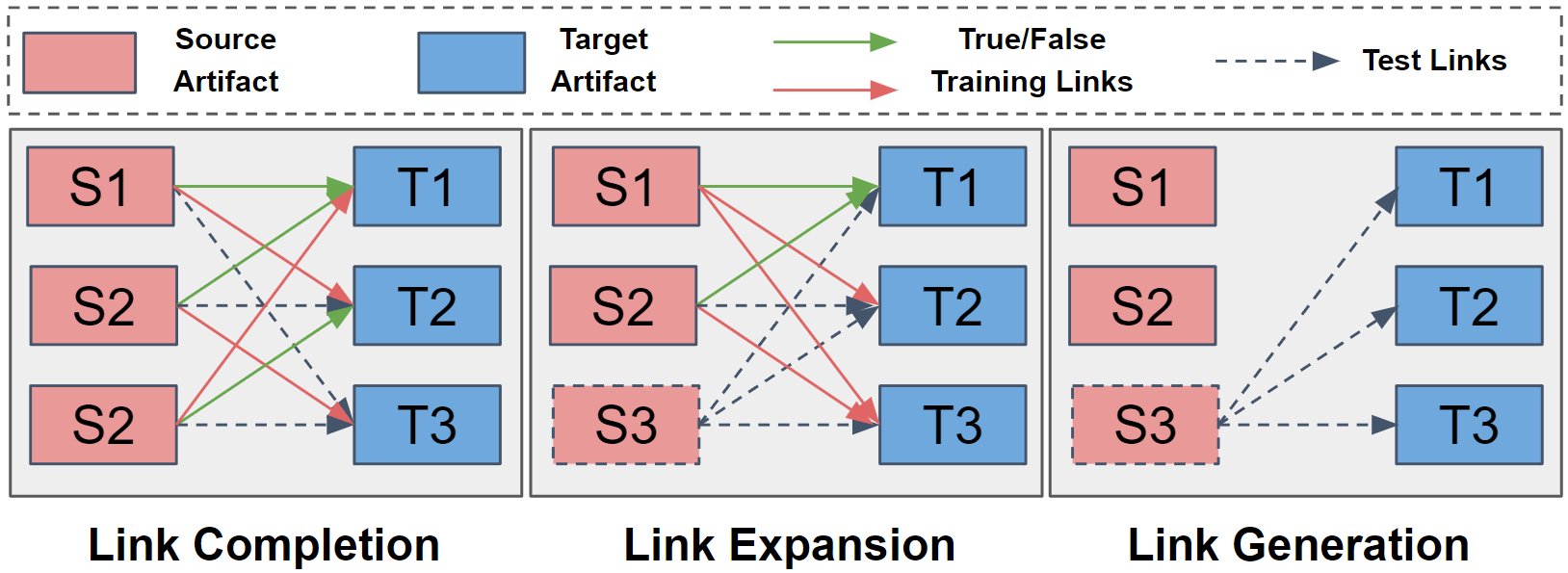}
    \caption{The data organization for experiments related to the three tracing tasks of  trace link completion (TLC), expansion (TLX) and generation (TLG) are supported though `split-by-link' and `split-by-artifact' strategies.}
    \label{tse_fig:trace_challenges}
\end{figure}

For the TLG generation task, we also adopted the `split-by-artifact' method to create the train/dev/test splits; however, we hid the trace links that were previously available as part of the training process. In the case of 0-shot (i.e., no training examples available), all links were masked; whereas in our experiment with 10-shot (i.e., 10 examples provided) we randomly selected 10 links from the training data, and allowed \model to use these links and their associated artifacts as examples. All other links were masked. The details of data splits are shown in Table.~\ref{tse_tab:data_detail} and the data split procedure is illustrated in Fig.~\ref{tse_fig:trace_challenges}.

To increase the reliability of our conclusion, we repeated each experiment five times using the train/validate/test folds created with different seeds, and our reported results reflect the average from the five runs. For the TLC, TLX and TLG experiments, we ran experiments on machines with one Quadro RTX 6000 GPU, whilst for the time consuming SE-Bert pretraining, we deployed our experiments on Azure servers using 8 Tesla-100V GPUs. 

\subsection{Evaluation Metrics}
We used F2 and Mean Average Precision (MAP) as our evaluation metrics. The F2 score measures the weighted harmonic mean of precision and recall  whilst favoring recall. We selected F2 over F1 because it is commonly used in software traceability experiments, where missing a link is more expensive than including a false link \cite{Hayes:AdvLinkGen}. 
\begin{align}
    F2 &= 5\cdot \frac{precision \cdot{recall} } {4 \cdot precision + recall}
\end{align}

MAP measures the overall ranking of the true links among all generated links. Each source artifact is treated as a query, and the Average Precision (AP) is calculated according to the ranking of its true links. In Eq.~\ref{lt_eq:ap}, $N_i$ refers to the number of true links for source artifact $i$, and Precision(j) calculates the precision for $link_j$ by processing the links ranked above it. It is computed as follows:
\begin{align}
    \label{lt_eq:ap}
    AP_i &= \frac{1}{N_i} \sum_{j=1}^{N} Precision(j)
\end{align}
MAP is then computed as the mean of all AP values as shown in Eq.~\ref{lt_eq:map}, where M refers to the number of source artifacts.  
\begin{align}
    \label{lt_eq:map}
    MAP &= \frac{1}{M} \sum_{i=1}^{M} AP_i
\end{align}

\section{Results And Discussion}
\label{tse_sec:results}
We now address the four research questions. For each question we describe the experiments that were conducted, analyze the results, and summarize the findings in a series of nine key observations.

\subsection{RQ1: How well does {\it \model} perform without the benefit of domain-specific transfer learning, and does it outperform classical IR trace models and other previously described DL tracing models?}
\label{tse_sec:rq1}
\begin{table*}[t!]
 \centering
\caption{Performance of BERT Classification Model and baselines on TLC, TLX and TLG without the benefit of the external knowledge sources and associated transfer learning techniques.  }
 \begin{subtable}[h]{\textwidth}
	    \centering
    \begin{tabular}{c|cc|cc|cc|cc|cc|cc} \toprule \hline
    & && \multicolumn{2}{c|}{CCHIT} & \multicolumn{2}{c|}{PTC } & \multicolumn{2}{c|}{Drone} & \multicolumn{2}{c|}{CM1} & \multicolumn{2}{c}{Avg Improve} \\
    & && F2 & MAP & F2 & MAP & F2 & MAP & F2 & MAP & F2 & MAP \\ \hline
    \multirow{9}{*}{ TLC } 
    & VSM &\Circle& 0.166 & 0.314 & 0.209 & 0.510 & 0.568 & 0.807 & 0.469 & 0.637 & - & - \\
    & LDA &\Circle& 0.069 & 0.119 & 0.090 & 0.328 & 0.460 & 0.593 & 0.409 & 0.466 & -57.50\% & -62.96\% \\
    & LSI &\Circle& 0.033 & 0.092 & 0.077 & 0.318 & 0.348 & 0.640 & 0.290 & 0.514 & -83.32\% & -65.95\% \\ 
    & TraceNN &\RIGHTcircle& 0.243 & 0.333 & 0.090 & 0.315 & 0.275 & 0.397 & 0.224 & 0.224 & -12.43\% & -34.47\%\\
    & DeepMatcher&\Circle & 0.137 & 0.260 & 0.066 & 0.390 & 0.556 & 0.725 & 0.294 & 0.470 & -37.60\% & -26.31\% \\ 
    & Bert &\Circle& \nmark{0.599} & 0.610 & 0.404 & \nmark{0.703} & 0.677 & 0.864 & \nmark{0.503} & \nmark{0.628}  & \nmark{186.70\%} & \nmark{72.85\%} \\
    & RoBerta&\Circle & 0.581 & 0.599 & \nmark{0.412} & 0.640 & \nmark{0.699} & \nmark{0.925} & 0.287 & 0.367  & 170.69\% & 59.16\% \\
    & XLNet &\Circle& 0.573 & 0.587 & 0.268 & 0.514 & 0.687 & 0.869 & 0.259 & 0.283 & 148.25\% & 45.39\%\\
    & DistillBert&\Circle & 0.568 & \nmark{0.633} & 0.345 & 0.641 & 0.651 & 0.904 & 0.264 & 0.288 & 154.52\% & 62.41\% \\ 
    \hline \bottomrule
    \multirow{9}{*}{ TLX } 
    & VSM &\Circle& 0.166 & 0.162 & 0.217 & 0.277 & 0.562 & 0.698 & 0.458 & 0.531  & - & -  \\
    & LDA &\Circle& 0.076 & 0.044 & 0.103 & 0.145 & 0.424 & 0.409 & 0.403 & 0.307 & -35.90\% & -51.04\% \\
    & LSI &\Circle& 0.071 & 0.032 & 0.081 & 0.140 & 0.408 & 0.502 & 0.295 & 0.376  & -45.79\% & -46.69\% \\ 
    & TraceNN &\RIGHTcircle& 0.086 & 0.093 & 0.092 & 0.097 & 0.056 & 0.027 & 0.177 & 0.124 & -64.33\% & -70.01\% \\
    & DeepMatcher&\Circle & 0.086 & 0.095 & 0.077 & 0.180 & 0.273 & 0.348 & 0.410 & 0.295 & -43.63\% & -42.72\%\\ 
    & Bert &\Circle& 0.284 & 0.211 & \nmark{0.488} & 0.580 & 0.541 & 0.716 & 0.318 & 0.258 & 40.37\% & 22.65\% \\
    & RoBerta &\Circle& \nmark{0.331} & 0.223 & 0.474 & \nmark{0.588} & \nmark{0.624} & \nmark{0.729} & 0.342 & \nmark{0.366} & \nmark{50.94\%} & \nmark{30.82\%}\\
    & XLNet &\Circle& 0.257 & 0.178 & 0.320 & 0.425 & 0.499 & 0.655 & \nmark{0.386} & 0.333 & 18.76\% & 4.98\%  \\
    & DistillBert &\Circle& 0.224 & \nmark{0.235} & 0.365 & 0.484 & 0.525 & 0.660 & 0.255 & 0.194 & 13.04\% & 12.75\% \\
    \hline
    TLG 
    & Bert &\Circle& 0.083 & 0.058 & 0.044 & 0.080 & 0.191 & 0.198 & 0.311 & 0.173 & -57.27\% & -78.53\% \\ \hline \bottomrule
    \end{tabular}
    \end{subtable}
    \Circle=Baseline, \RIGHTcircle=Proposed architecture without the benefit of transfer learning

    \label{tse_tab:rq1}
\end{table*}

To answer this question we evaluated \model using four general purpose LMs and three different classical IR tracing models.  The general purpose LMs were  `bert-base-uncased', `roberta-base', `xlnet-base-cased' and `distilbert-base-uncased', while the classical IR models were VSM, LDA and LSI. In addition, we also evaluated TraceNN  \cite{guo2017semantically} as discussed in Section \ref{tse_sec:related_work}, and DeepMatcher \cite{haering2021automatically}.

For TraceNN, we implemented the model according to the authors' description. In their study, RNNs with LSTM and BiGRU architectures are discussed and compared. We chose the BiGRU version as it achieved better tracing results in their study and also outperformed LSTM when used in our own prior work  \cite{ICSE21-Jinfeng} to trace from text to code. We applied the default configurations specified in their paper, except we adjusted the neural size and training epochs to adapt the model to our experiment projects' size. In addition, we added a dropout, with a rate of 0.2, to the MLP layers to reduce the chance of overfitting. Finally, we applied the widely used pre-built Glove word embeddings  \cite{pennington2014glove} to initialize the embedding layer for TraceNN. 

DeepMatcher is an LM based document encoder which transforms artifacts into a vector representation and then uses Cosine similarity to calculate  relevance between pairs of vectors. DeepMatcher uses DistillBert to generate token embeddings for the tokens in noun phrases, and then takes the average of those vectors to produce the final artifact representation. We followed the authors' description to implement this model and used the same libraries mentioned in their work. 

Results are shown in Table~\ref{tse_tab:rq1} for TLC and TLX. TLG is identical to TLX for techniques that do not have any inherent training (e.g., VSM, LDA, LSI, and DeepMatcher). TraceNN explicitly requires training data, and cannot be executed without it. Bert, RoBerta, XLNet, and DistillBert all leverage trace links as part of their fine-tuning step, and cannot be expected to perform well without links.  To demonstrate this, we report the use of Bert as a representative LM model applied to LTG in the final row of Table \ref{tse_tab:rq1}.

The two rightmost columns in Table ~\ref{tse_tab:rq1}  compare the performance for each approach averaged across each of the four projects versus the average performance of VSM.  These results show that VSM outperformed the topic modeling approaches of LDA and LSI, as reported consistently in prior work (e.g., \cite{lin2022information,lohar2013improving}). It also outperformed DeepMatcher for TLC, TLX, and TLG, and outperformed TraceNN for the two tasks that TraceNN was applied to (i.e., TLX and TLC). 

Despite promising previously published results \cite{guo2017semantically}, we observed that TraceNN only outperformed VSM in one out of eight cases, namely the TLC task in the CCHIT dataset. The learning curve indicated that TraceNN quickly adapted to the training data with a rapid drop in training loss, however the validation loss converged at a relatively high value. Guo \etal described this overfitting phenomenon as the `glass ceiling', which is the major impedance of the RNN based trace model. We also observed that TraceNN did not converge for the TLX problem, most likely due to the gap between training and testing data in the TLX expansion problem, where the model is asked to generate links for artifacts that have never been seen in training. DeepMatcher achieved higher MAP scores but lower F2 scores than TraceNN, and also did not outperform VSM on either the TLC or TLX tracing tasks.

The general LM based models (i.e., Bert, RoBerta, XLNet, and DistillBert) all achieved significantly better performance across the four projects than the classical IR approaches, TraceNN, and DeepMatcher. The Bert based model improved over VSM for both the TLC and TLX tasks by an average of 186.70\% and 72.85\% for F2 and MAP respectively in the TLC task, and by 40.37\% (F2) and 22.65\% (MAP) for TLX. The RoBerta model outperformed Bert for the TLX task achieving an additional 10\% (F2) and 8\% (MAP). We believe that the additional corpus used by RoBerta provided the extra knowledge needed to mitigate the terminology gap between training and test data in order to generate links for previously unseen data. This contrasted with the TLC task in which the training data was able to provide better supervision than the knowledge extracted from the additional corpus. We hypothesized that the knowledge from RoBerta's additional pretraining may have conflicted with the fine-tuning procedure, because the additional corpus used by RoBerta was not related to our target project domain. 

To support our observation that LM based models do not generally perform well on the TLG tracing task, we also report the Bert results without the benefit of any training, and observe that Bert is outperformed by VSM by 57.27\% (F2) and 78.53\% (MAP) respectively. This poor performance is because Bert is not calibrated to the project data and therefore tends to produce somewhat haphazard results. These findings are summarized through the following observation.\vspace{2pt}

\noindent \mybox{lightgray}{{{\bf Observation \#1:~} Currently available \pretrained language models (LMs) do not perform well on tracing tasks without the benefit of transfer learning.}}\vspace{3pt}\\

\subsection{RQ2: Which, if any, of the three transfer learning strategies, produce LMs that outperform the original general-purpose LMs with respect to the TLC and TLX tracing tasks?}
In this section we address RQ2 through evaluating the effectiveness of each of the three transfer learning strategies, as previously described in Section \ref{sec:transfer_strategies}, applied to the TLC and TLX tracing tasks. We do not include TLG in this evaluation, as additional transfer learning strategies cannot overcome the initial lack of finetuning discussed in RQ1. For each strategy we compare our novel approaches (labeled \CIRCLE) against a set or related baselines (labeled \Circle).\vspace{3pt}

\begin{table*}[t!]
	\centering
	\caption{Accuracy achieved by pretraining the LM on Git Corpus and Project Corpus for the classification model. Average improvement results (right hand side) represent comparisons to Bert (top row).}
	    \begin{subtable}[h]{\textwidth}
	    \centering
        \begin{tabular}{cc|cc|cc|cc|cc|cc|cc}
        \toprule
        \hline
        {\bf
        Completion}& & Arch&Trans & \multicolumn{2}{c|}{CCHIT} & \multicolumn{2}{c|}{PTC } & \multicolumn{2}{c|}{Drone} & \multicolumn{2}{c|}{CM1} & \multicolumn{2}{c}{Avg Improve} \\
        (TLC)& &Style & Learn & F2 & MAP & F2 & MAP & F2 & MAP & F2 & MAP & F2 & MAP \\ \hline
        Bert&\Circle &CLS&n/a & 0.599 & 0.610 & 0.404 & 0.703 & 0.677 & 0.864 & 0.503 & 0.628 & - & - \\ 
        Roberta&\Circle &CLS&n/a  & 0.581 & 0.599 & 0.412 & 0.640 & 0.699 & 0.925 & 0.287 & 0.367 & -10.15\% & -11.29\% \\
        SciBert& \Circle&CLS &\PRE & 0.602 & \nmark{0.664} & \nmark{0.441} & \nmark{0.740} & 0.683 & 0.904 & 0.578 & 0.705 & 6.33\% & 7.74\% \\ \hline
        \clssebert& \CIRCLE&CLS&\PRE & 0.612 & 0.654 & 0.389 & 0.634 & 0.714 & \nmark{0.943} & 0.375 & 0.564 & -5.43\% & -0.93\% \\
        \clsproj& \CIRCLE&CLS&\PAD & \nmark{0.616} & 0.641 & 0.412 & 0.692 & \nmark{0.715} & 0.918 & \nmark{0.592} & \nmark{0.772} & \nmark{6.96\%} & \nmark{8.18\%} \\ 
        
            \toprule
            \hline
            {\bf Expansion}& & Arch&Trans & \multicolumn{2}{c|}{CCHIT} & \multicolumn{2}{c|}{PTC } & \multicolumn{2}{c|}{Drone} & \multicolumn{2}{c|}{CM1} & \multicolumn{2}{c}{Avg Improve} \\
            (TLX)&& Style & Learn & F2 & MAP & F2 & MAP & F2 & MAP & F2 & MAP & F2 & MAP \\ \hline
            Bert &\Circle&CLS&n/a  & 0.284 & 0.211 & 0.488 & 0.580 & 0.541 & 0.716 & 0.318 & 0.258 & - & - \\ 
            Roberta& \Circle&CLS&n/a  & 0.331 & 0.223 & 0.474 & 0.588 & 0.624 & 0.729 & 0.342 & 0.366 & 9.19\% & 12.68\% \\
            SciBert&\Circle &CLS &\PRE  & 0.384 & 0.295 & \nmark{0.493} & \nmark{0.634} & 0.589 & 0.713 & \nmark{0.512} & 0.409 & \nmark{26.65\%} & 26.81\% \\ \hline
            \clssebert &\CIRCLE&CLS&\PRE & 0.346 & 0.264 & 0.456 & 0.576 & \nmark{0.628} & \nmark{0.789} & 0.468 & \nmark{0.460} & 19.68\% & \nmark{28.29\%} \\
            \clsproj &\CIRCLE&CLS&\PAD & 0.371 & 0.253 & 0.479 & 0.580 & 0.608 & 0.747 & 0.479 & 0.394 & 23.01\% & 19.32\% \\  \hline
            \end{tabular}
        \end{subtable}
               \begin{tablenotes}
            \item \centering \modelExplain, \ADJ* uses general NLP tasks as adjacent task
             \item \Circle=Baseline, \CIRCLE=Proposed transfer learning technique
        \end{tablenotes}
        \label{tse_tab:pretrain_transfer}
\end{table*}
\noindent{\it Pretraining Transfer Strategies:~} First, we explored how the two pretraining based transfer learning strategies, named `DomainLM' (cf.~Section \ref{sec:transfer_strategies} and `Target Project Adaptation' (cf.~Section \ref{tse_sec:tapt}, impact the trace performance. Results are reported in  Table.~\ref{tse_tab:pretrain_transfer}.
As previously explained, \clssebert and \clsproj both represent LM based classification trace models that deploy our proposed transfer learning strategies. We used the Bert based model as a baseline given its strong performance in the RQ1 experiments. We also added SciBert \cite{beltagy2019scibert} for additional comparison purposes. SciBert is a Bert model \pretrained to perform NLP tasks in general scientific areas. Its corpus contains 1.14M papers, including 18\% from the computer science domain and 82\% from the biomedical domain, and as such is a reasonable match for our four technical domains of health-care, train controls, UAV, and space exploration (CM1) 

The results in this table show that \clsproj effectively improved the trace performance for both the TLC and TLX tasks and achieved an average improvement of 6.96\% (F2) and 8.18\% (MAP) over the Bert baseline for the TLC task, outperforming the SciBert model. For the TLX task, Proj-CLS achieved 23.01\% (F2) and 19.32\% (MAP) improvements, but did not outperform SciBert. These results suggest the following:\vspace{2pt}

\noindent \mybox{lightgray}{{{\bf Observation \#2:~} Pretraining using a project corpus collected through our data pipeline can effectively improve the accuracy of TLX and TLC tracing tasks in the associated project.}}\vspace{3pt}\\

\begin{table*}[t!]
	\centering
	\caption{Accuracy eciehved through the use of transfer learning from adjacent tasks. Average improvement results (right hand side) represent comparisons to Bert (top row).}
	    \begin{subtable}[h]{\textwidth}
	    	\centering
            \begin{tabular}{cc|cl|cc|cc|cc|cc|cc} \toprule \hline
            Completion &&  Arch&Trans &\multicolumn{2}{c|}{CCHIT} & \multicolumn{2}{c|}{PTC } & \multicolumn{2}{c|}{Drone} & \multicolumn{2}{c|}{CM1} & \multicolumn{2}{c}{Avg Improve} \\ 
            && Style&Learn&F2 & MAP & F2 & MAP & F2 & MAP & F2 & MAP & F2 & MAP \\ \hline
            Bert&\Circle & CLS &n/a & 0.599 & 0.610 & 0.404 & 0.703 & 0.677 & 0.864 & 0.503 & 0.628 & - & - \\
            SciBert&\Circle & CLS & \PRE  & 0.602 & 0.664 & \nmark{0.441} & \nmark{0.740} & 0.683 & 0.904 & 0.578 & 0.705 & 6.33\% & 7.74\% \\ 
            \clstask&\CIRCLE & CLS & \ADJ & 0.610 & \nmark{0.679} & 0.439 & 0.733 & 0.718 & \nmark{0.948} & \nmark{0.626} & \nmark{0.817} & \nmark{10.21\%} & \nmark{13.84\%} \\ \hline
            MNLI &\Circle& CLS & \ADJ*  & 0.508 & 0.558 & 0.439 & 0.657 & \nmark{0.723} & 0.927 & 0.456 & 0.549 & -2.30\% & -5.07\% \\
            MRPC &\Circle & CLS & \ADJ*& 0.598 & 0.651 & 0.428 & 0.737 & 0.699 & 0.869 & 0.537 & 0.673 & 3.91\% & 4.83\% \\
            QNLI &\Circle& CLS & \ADJ* & 0.591 & 0.643 & 0.389 & 0.691 & 0.711 & 0.867 & 0.318 & 0.362 & -9.22\% & -9.54\% \\
            RTE & \Circle&CLS & \ADJ* & 0.602 & 0.633 & 0.403 & 0.660 & 0.635 & 0.871 & 0.251 & 0.335 & -13.98\% & -12.02\% \\
            STS-B&\Circle & CLS & \ADJ* & \nmark{0.623} & 0.628 & 0.422 & 0.624 & 0.714 & 0.917 & 0.564 & 0.746 & 6.55\% & 4.20\% \\ 
            \end{tabular}
        \end{subtable}
        
        \begin{subtable}[h]{\textwidth}
            \centering
            \begin{tabular}{cc|cl|cc|cc|cc|cc|cc}
            \toprule
            \hline
            Expansion &&  Arch&Trans &\multicolumn{2}{c|}{CCHIT} & \multicolumn{2}{c|}{PTC } & \multicolumn{2}{c|}{Drone} & \multicolumn{2}{c|}{CM1} & \multicolumn{2}{c}{Avg Improve} \\ 
            && Style&Learn& F2 & MAP & F2 & MAP & F2 & MAP & F2 & MAP & F2 & MAP \\ \hline
            Bert &\Circle& CLS &n/a  & 0.284 & 0.211 & 0.488 & 0.580 & 0.541 & 0.716 & 0.318 & 0.258 & - & - \\
            SciBert&\Circle & CLS & \PRE  & 0.384 & 0.295 & \nmark{0.493} & \nmark{0.634} & 0.589 & 0.713 & 0.512 & 0.409 & 26.65\% & 26.81\% \\ 
            \clstask&\CIRCLE & CLS & \ADJ & 0.392 & \nmark{0.317} & 0.476 & 0.588 & \nmark{0.611} & 0.743 & \nmark{0.515} & 0.423 & \nmark{27.69\%} & \nmark{29.87\%} \\ \hline
            MNLI &\Circle& CLS & \ADJ*  & 0.333 & 0.246 & 0.472 & 0.566 & 0.550 & 0.663 & 0.371 & 0.310 & 8.15\% & 6.74\% \\
            MRPC &\Circle& CLS & \ADJ*  & 0.364 & 0.271 & 0.486 & 0.595 & 0.610 & \nmark{0.764} & 0.442 & 0.382 & 19.89\% & 21.43\% \\
            QNLI &\Circle& CLS & \ADJ*  & 0.380 & 0.266 & 0.438 & 0.546 & 0.557 & 0.670 & 0.276 & 0.230 & 3.37\% & 0.81\% \\
            RTE &\Circle& CLS & \ADJ*  & 0.379 & 0.260 & 0.437 & 0.531 & 0.542 & 0.678 & 0.294 & 0.202 & 4.00\% & -2.94\% \\
            STS-B &\Circle& CLS & \ADJ*  & \nmark{0.393} & 0.269 & 0.450 & 0.552 & 0.578 & 0.726 & 0.448 & \nmark{0.427} & 19.65\% & 22.45\% \\ \hline
            \end{tabular}
        \end{subtable}
        \begin{tablenotes}
            \item \centering \modelExplain, \ADJ* uses general NLP tasks as adjacent task
            \item \Circle=Baseline, \CIRCLE=Proposed transfer learning technique
        \end{tablenotes}
 
        \label{tse_tab:adj_task}
\end{table*}
\begin{table*}[t!]
	\centering
	\caption{Performance of CSE models with transfer learning}
	    \begin{subtable}[h]{\textwidth}
	    \centering
        \begin{tabular}{cc|cl|cc|cc|cc|cc|cc}\toprule \hline
        Completion && Arch&Trans & \multicolumn{2}{c|}{CCHIT} & \multicolumn{2}{c|}{PTC } & \multicolumn{2}{c|}{Drone} & \multicolumn{2}{c|}{CM1} & \multicolumn{2}{c}{Avg Improve}\\
        && Style&Learn & F2 & MAP & F2 & MAP & F2 & MAP & F2 &MAP & F2 & MAP \\ \hline
        Bert&\Circle& CLS& n/a & \nmark{0.599} & \nmark{0.610} & \nmark{0.404} & 0.703 & 0.677 & 0.864 & 0.503 & 0.628 & - & - \\
        \rankglue &\Circle& CSE & \ADJ* & 0.334 & 0.512 & 0.283 & 0.778 & \nmark{0.701} & 0.914 & 0.518 & \nmark{0.713} & -16.95\% & -5.49\% \\
        \ranktask &\CIRCLE& CSE & \ADJ & 0.396 & 0.586 & 0.312 & \nmark{0.820} & 0.681 & \nmark{0.954} & 0.511 & \nmark{0.713} & -13.65\% & 0.00\% \\
        \rankproj & \CIRCLE&CSE & \PAD & 0.198 & 0.334 & 0.173 & 0.510 & 0.535 & 0.859 & \nmark{0.548} & 0.674 & -34.05\% & -24.03\% \\

        
        \toprule \hline 
        Expansion && Arch&Trans & \multicolumn{2}{c|}{CCHIT} & \multicolumn{2}{c|}{PTC } & \multicolumn{2}{c|}{Drone} & \multicolumn{2}{c|}{CM1} & \multicolumn{2}{c}{Avg Improve}\\
        & &Style&Learn & F2 & MAP & F2 & MAP & F2 & MAP & F2 & MAP & F2 & MAP \\ \hline
        Bert &\Circle& CSE & n/a & 0.284 & 0.211 & \nmark{0.488} & \nmark{0.580} & 0.541 & 0.716 & 0.318 & 0.258 & - & - \\
        \rankglue &\Circle& CSE & \ADJ* & 0.311 & 0.237 & 0.309 & 0.506 & \nmark{0.611} & 0.812 & 0.518 & 0.473 & 12.23\% & 24.20\% \\
        \ranktask& \CIRCLE&CSE & \ADJ & \nmark{0.336} & \nmark{0.289} & 0.345 & 0.552 & 0.592 & \nmark{0.829} & 0.504 & \nmark{0.583} & \nmark{14.33\%} & \nmark{43.55\%} \\
        \rankproj & \CIRCLE&CSE & \PAD & 0.228 & 0.203 & 0.163 & 0.233 & 0.510 & 0.672 & \nmark{0.520} & 0.492 & -7.02\% & 5.26\% \\ \hline
            \end{tabular}
        \end{subtable}
        \begin{tablenotes}
            \item \centering \modelExplain, \ADJ* uses general NLP tasks as adjacent task
            \item \Circle=Baseline, \CIRCLE=Proposed transfer learning technique
        \end{tablenotes}
        \label{tse_tab:cse}
\end{table*}
The performance of \clssebert was somewhat mixed. It achieved its best average MAP on the TLX task, but performed worse than the Bert baseline on the TLC problem. Looking at individual projects, we see that for TLC tasks, \clssebert underperformed on the  PTC and CM1 projects but achieved  the best and moderate results on Drone and CCHIT projects respectively. These results can be explained by the availability, or lack of availability, of relevant projects in the Git corpus with insufficient coverage for the two underperforming domains. As an OSS community, Github has fewer repositories related to space instruments (CM1), and train control (PTC), whilst having 124 repositories \footnote{\href{https://github.com/search?p=1&q=open\%2Behr\&type=Repositories}{EHR repos}} related to EHR (Electronic Health Records), and 33k repositories related to drones \footnote{\href{https://github.com/search?q=drone\&type=Repositories}{Drone Repos}}. 
Although SE-BERT did not perform as well as the SciBERT and \clsproj models in this study, it could potentially be useful to the OSS Engineering community as a \pretrained LM dedicated to the SE domain.  We therefore publicly release \clssebert to support future research in NL tasks within the  OSS domain. 
\vspace{2pt}

\noindent \mybox{lightgray}{{{\bf Observation \#3:~} OSS supported pretraining of LMs is only effective when a sufficiently rich corpus of relevant OSS projects is available for the project domain.}}\vspace{3pt}\\

\noindent{\it Pretraining Transfer Strategies:~}Our second experiment focused on task-level transfer for CLS, with results reported in Table.~\ref{tse_tab:adj_task}. In addition to the Bert and SciBert baselines, we also include models previously developed for general NL-NL tasks and provided by the GLUE dataset \cite{wang2018glue}. By comparing the performance of \clstask and these models, we were able to evaluate whether Commit-Issue trace links provide a better knowledge source than other NL-NL tasks that have been shown to perform well for other more general NLP tasks. Descriptions of the  general NLP tasks that we included in our experiment are provided in Table~\ref{tse_tab:nlp_tasks}. While  many different training tasks have been explored in prior NLP research, we selected these five tasks because they focused on classifying sentence-to-sentence relationships through exploiting token level associations. The tasks, which include predicting entailments and contradictions, evaluating paraphrases, determining whether a sentence provides an answer to a specific question, and checking for semantic relatedness between two sentences, were chosen because they are similar in nature to the tracing tasks and are therefore more likely to be effective for supporting transfer learning for TLX, TLC, and TLG related tasks. 
\begin{table}[!tb]
	\centering
    \caption{Extra text-2-text tasks from GLUE dataset for improving \model performance at fine-tuning stage}
    \addtolength{\tabcolsep}{-2pt}
    \begin{tabular}{|c|p{2.6cm}|p{4.7cm}|}
     \hline
        \textbf{Task}& \textbf{\hfil Name} &\hfil \textbf{Description }\\ \hline
        MNLI & Multi-Genre Natural Language Inference & Predicts whether S1 is entailed, neutral or in contradiction. to S2\\ \hline
        MRPC &  Microsoft Research Paraphrase & Determines whether S2 paraphrases S1 without changing its meaning. \\ \hline
        QNLI & Question Natural Language Inference & Determines whether S1 contains an answer to question S2. \\ \hline
        RTE & Recognizing Textual Entailment & Binary classification task predicting whether S2 entails  S1.  \\ \hline
        STS-B & Sentence Semantic Similarity& Evaluates whether two sentences are semantically related. \\
        \hline
    \end{tabular}
    \vspace{5pt}\\
    S1 = First sentence; S2 = Second sentence
    \label{tse_tab:nlp_tasks}
\end{table} 

Results are reported in Table.~\ref{tse_tab:adj_task}. 
Among all the general NL-NL tasks, MPRC and STS-B achieved the greatest improvements over the BERT baseline for the TLC and TLX tasks, with  MRPC improving performance by 3.91\% (F2) and 4.83\% (MAP) for TLC, and by a more significant amount of 19.89\% (F2) and 21.43\% (MAP) for TLX. Similarly, STS-B improved TLC by  6.55\% (F2) and 4.2\% (MAP) for TLC, and by 19.65\% (F2) and 22.45\% (MAP) for TLX. Intuitively, the reason that MRPC and STS-B outperformed the other training tasks, was that  their focus on determining similarity between two sentences, which more closely matched the objective of software traceability. However, the \clstask model, which applied the classification task to the Issue and Commit Git links instead of applying it to more general pairs of sentences, performed even better, achieving 10.21\% (F2) and 13.84\% (MAP) improvement for TLC, and 27.69\% (F2) and 29.87\% (MAP) improvement on TLX. Finally, the \clstask also outperformed the SciBert based trace model.\vspace{2pt}

\noindent \mybox{lightgray}{{{\bf Observation \#4:~} General NL tasks focused on detecting similarity between sentence pairs supported transfer learning for the TLC and TLX tracing tasks; however, the \clstask approach, which applied these tasks to issue-commit links, performed even better.}}\vspace{3pt}\\

\noindent{\it Transfer Learning applied to CSE Architectures:~}Finally, we also explored transfer learning applied to the CSE architecture. The authors of SimCSE \cite{gao2021simcse}, released their model trained using task-level knowledge transfer on the GLUE datasets, and reported that it  outperformed its counterpart, the  Bert-RANK model, on several NL-NL NLP tasks. Therefore, we integrated their approach into our own CSE architecture, naming it GLUE-RANK and adopting it as our RANK model baseline. Further, we compared it against our own Bert-CLS model (described in Section \ref{tse_sec:rq1}) as a CLS baseline. 

As reported in Table~\ref{tse_tab:cse}, for the TLC task, none of the CSE based models outperformed Bert-CLS; however, for the TLX task both \ranktask and \rankglue significantly outperformed Bert-CLS.  \ranktask achieved a 43.55\% improvement in MAP score for LTX, which is significantly higher than the 29.87\% improvement in MAP achieved by \ranktask. These results suggest that the CSE architecture has a stronger generalization ability than CLS and tends to perform better in cases where the train and test data have a larger distribution gap; whilst having a relatively weaker ability to fit the training data than the CLS architecture.

In summary, we observed that all three types of transfer learning strategies were beneficial in some way for improving tracing performance. The datasets we collected, including the Git corpus and Git links, as well as the Project corpus, were generally more effective as knowledge sources than their more general NLP counterparts.  \vspace{2pt}

\noindent \mybox{lightgray}{{{\bf Observation \#5:~} Task-related knowledge transfer returned marked improvements in trace accuracy, especially when training tasks were performed using sentence matching tasks trained on the artifacts connected by issue-commit links.}}\vspace{3pt}\\

Among the three strategies that we explored, task-level transfer achieved the overall best performance evidenced by the fact that \clstask and \ranktask outperformed other variants for each of the tasks.\vspace{2pt}

\noindent \mybox{lightgray}{{{\bf Observation \#6:~} Task-level transfer was more effective than other transfer strategies that used domain-related pretraining and adaptation of the LM.}}\vspace{3pt}\\

\subsection{RQ3: Can LM models outperform classical IR methods on the TLG task when a small number of training examples are provided?}
Our previous experiments focused on TLC and TLX tasks for which a training set of links was available; however, in this research question, we explored the TLG task for which no training links were available (i.e., 0-shot). We also investigated the potential improvement of providing 10 example trace links (i.e., 10-shot) for training purposes. The analysis of more varied numbers of training links are left for future work. 

For models, such as VSM, which tune their parameters based on text only, we gave access to all source and target artifacts in the training dataset to conduct indexing and self-supervised training. Based on the results of RQ2, we focused on the \clstask and \ranktask variants as they achieved the best results for TLC and TLX experiments. For comparison purposes, we also included VSM, SciBert-CLS and \rankglue as representative baselines for information retrieval, classification, and contrastive sentence embedding (CSE) techniques respectively. These were selected because of their superior performance in our previous experiments.
\begin{table*}[t!]
	\centering
	\caption{Performance of CLS and CSE models on 0 shot and 10 shots link generation problems. Average improvement results (right hand side) represent comparisons to VSM (top row).}
	    \begin{subtable}[h]{\textwidth}
	    \centering
	    \begin{tabular}{ cc|cl|cc|cc|cc|cc|cc}
	    \toprule \hline
        Generation-0&& Arch&Trans&\multicolumn{2}{c|}{CCHIT} & \multicolumn{2}{c|}{PTC } & \multicolumn{2}{c|}{Drone} & \multicolumn{2}{c|}{CM1} & \multicolumn{2}{c}{Avg Improve} \\
        (TLG)&& Style&Learn&F2 & MAP & F2 & MAP & F2 & MAP & F2 & MAP & F2 & MAP \\ \hline
        VSM & \Circle & CSE &n/a&0.166 & 0.162 & \nmark{0.217} & \nmark{0.277} & \nmark{0.562} & \nmark{0.698} & 0.458 & 0.531 & - & - \\
        SciBert-CLS & \Circle & CLS&PRE&0.099 & 0.101 & 0.034 & 0.065 & 0.108 & 0.077 & 0.280 & 0.175 & -61.00\% & -67.52\% \\
        \rankglue & \Circle &CSE&ADJ*& 0.177 & 0.202 & 0.113 & 0.242 & 0.324 & 0.502 & 0.593 & 0.543 & -13.49\% & -3.35\% \\ \hline
        \clstask &\CIRCLE &CLS&ADJ& \nmark{0.335} & \nmark{0.229} & 0.183 & 0.219 & 0.493 & 0.312 & \nmark{0.605} & \nmark{0.564} & \nmark{26.59\%} & -7.08\% \\
        \ranktask &\CIRCLE &CSE&ADJ & 0.228 & 0.203 & 0.163 & 0.233 & 0.510 & 0.672 & 0.520 & 0.492 & 4.33\% & -0.34\% \\

            \toprule \hline
            Generation-10 && \multicolumn{2}{c}{} &\multicolumn{2}{c}{} & \multicolumn{2}{c}{ } & \multicolumn{2}{c}{} & \multicolumn{2}{c}{} & \multicolumn{2}{c}{} \\ \hline
            SciBert-CLS & \Circle & CLS & \PRE & 0.182 & 0.075 & 0.065 & 0.140 & 0.355 & 0.428 & 0.310 & 0.239 & -32.32\% & -49.22\% \\
            \rankglue & \Circle   & CSE & \ADJ* & 0.259 & 0.150 & \nmark{0.198} & \nmark{0.332} & 0.491 & 0.692 & \nmark{0.511} & 0.488 & 11.69\% & 0.98\% \\ \hline
            \clstask &\CIRCLE   & CLS & \ADJ & \nmark{0.337} & \nmark{0.250} & 0.183 & 0.273 & \nmark{0.532} & 0.603 & \nmark{0.604} & \nmark{0.562} & \nmark{28.58}\% & \nmark{11.39\%} \\
            \ranktask &\CIRCLE   & CSE & \ADJ & 0.276 & 0.209 & 0.149 & 0.253 & 0.508 & 0.688 & 0.475 & 0.403 & 7.28\% & -1.31\% \\ \hline
            \end{tabular}
        \end{subtable}
        \begin{tablenotes}
            \item \centering \modelExplain, \ADJ* uses general NLP tasks as adjacent task
        \end{tablenotes}
        \label{tse_tab:generation}
\end{table*}

Results are reported in Table.~\ref{tse_tab:generation}. In the 0-shot experiments, the classical VSM model outperformed both  SciBert-CLS and \rankglue. SciBert-CLS underperformed because it needs training data to tune its classification network. For \rankglue, whilst we allowed it to use the raw artifacts (without any links) to conduct self-supervised learning, the self-supervision signal in the ranking task was unable to compete with VSM's results. 

The \clstask however outperformed VSM by 26.59\% with respect to average F2 but exhibited a loss in MAP of 7.08\%. While these results are slightly mixed, they suggest that the issue-commit links played a role in the task-based transfer by improving trace link accuracy even when no training links were available. The simultaneous gain in F2 but loss in MAP indicates that more of the targeted links achieved similarity scores above the prescribed threshold; but that they were not ranked sufficiently high in the ordered list to improve MAP.  The \ranktask model, which uses the Issue-Commit links as a knowledge source, also outperformed VSM by 4.33\% of F2 on average and further supports our findings. 

Results for the 10-shot showed that  both SciBert-CLS and \rankglue benefited from even a small set of training examples; however, average results for SciBert-CLS were still lower than VSM by approximately  32.32\% (F2) and 49.22\% (MAP). In contrast, \rankglue outperformed VSM by 11.69\% (F2) and 0.98\% (MAP). In the case of \rankproj, providing ten examples improved performance on CCHIT, Drone, and PTC, but not on CM1. In fact, the 10-shot results for CM1 were worse than the 0-shot results! This was likely due to the complexity of the domain as well as the underlying architecture. CSE architectures use a negative selection mechanism, meaning that when the CSE model creates in-batch negative examples, it pairs source and target artifacts within the training batch to create a pool of negative links. However, this approach can incorrectly label positive links as negative ones, introducing noise into the training data, and resulting in lowered performance. This effect is more likely to occur in small projects such as CMI.

In contrast, \clstask dramatically benefited from the ten training examples that were provided. The difference in F2 scores for \clstask versus VSM was approximately equivalent for 0-shot and 10-shot (i.e., 26.59\% vs. 28.58\%); however, with 0-shot, the difference in MAP scores for \clstask  versus VSM was negative (i.e., -7.08\%), while with 10-shot they improved over VSM by 11.39\%, making it ultimately the overall best performing approach for TLG tasks.  Furthermore, whereas SciBert-CSL and \clstask, were both able to achieve good results in resource rich tracing tasks such as TLC and TLX tasks, the \clstask model performed better on resource limited cases by leveraging similarities between the downstream and the and adjacent tasks. \vspace{2pt}

\noindent \mybox{lightgray}{{{\bf Observation \#7:~} Transfer learning techniques that leveraged adjacent tasks returned mixed results when applied to TLG. Results were best when it was used in conjunction with the classical CLS architecture, where F2 scores improved but MAP scores reduced in comparison to the VSM baseline. However, when even 10 training examples were provided, the combination of task-based transfer and the classical (CLS) architecture returned improvements over VSM. }}\vspace{3pt}\\

\subsection{RQ4: What is the overall best method for supporting all
three NL tracing tasks?}
Finally, based on observations from RQ1, we conclude that general LM based trace models outperform both conventional IR models and the RNN based model when applied to the TLC and TLX tracing tasks. Based on results from RQ2, all three transfer learning approaches evaluated in this study improved the performance of the general LM based trace model regardless of whether CLS or CSE architectures were used. However, the best performance was achieved when using adjacent tracing tasks within the CLS architecture, where performance for both TLC and TLX improved by more than 20\% for both F2 and MAP scores. In RQ3, we further evaluated \clstask and \ranktask, as the two best variants identified in RQ2, to explore their potential for supporting TLG in cases where links needed to be generated from scratch.  Our results indicated that the \clstask model was able to effectively use the knowledge learned from adjacent tasks when few links were available for training purposes. It outperformed VSM by 26.59\% F2 when no training examples were provided at all. However, when humans provided even ten links as examples, the model outperformed the VSM model with an increase of 28.58\% (F2) and 11.39\% (MAP). Its performance was better than the SciBert-CLS baseline, which was effective only in resource rich scenarios (i.e., TLC, TLX). These results suggest that \clstask is the overall best model in cases where a single model is desirable to support all three tracing tasks of TLC, TLE, and TLG.\vspace{3pt}

\noindent \mybox{lightgray}{{{\bf Observation \#8:~} The best overall performer across all three tracing tasks (i.e., TLC, TLX, TLG) is \clstask which utilizes transfer learning based on adjacent tasks (built using commit-issue links) within the context of a CLS architecture. }}\vspace{3pt}\\

We also performed preliminary experiments to combine the strategies into a single solution. we applied transfer strategies in a sequential order that matched the natural order of our pipeline by first preparing the SE-BERT language model and then conducting transfer learning on adjacent tasks.  
However, neither of the three two-way combinations, nor the three-way combination outperformed individual strategies in our experiments. We therefore leave further investigation of this open issue to future work as discussed in the concluding section of this paper.\vspace{3pt}

\noindent \mybox{lightgray}{{{\bf Observation \#9:~} Combining the three knowledge transfer techniques in a pipeline did not outperform the best individual approach. Exploring combinations of techniques is left as an open research question. }}\vspace{3pt}\\

\section{threats to validity }
\label{tse_sec:threat}
Our study is impacted by three primary threats to validity. 
First, due to the inclusion and exclusion criteria we established for our study, along with the low availability of industrial project datasets, we evaluated our approach against only four software projects. However, these were taken from four different domains with three of them representing industrial or government projects, and the other representing a large academic project deployed in the physical world and developed by a diverse team of academic and professional developers. Nevertheless, given this limitation we are not able to more fully generalize our findings across other domains or even across a broader set of projects with more diverse terminology, templates, or styles of requirements specifications. 

Second, although our experiment datasets are quite large in comparison to many previously published traceability papers that focus on traditional software engineering projects, the project sizes are still relatively small in comparison to many real-world software projects, and it is well known that accuracy of tracing results are impacted negatively by project size due to the `needle-in-the-haystack' phenomenon.  This introduces the risk that accuracy might be negatively impacted as the size of the project grows; however, based on our observations in this study, and the general behavior of deep learning techniques, we expect larger project sizes to actually be beneficial for the performance of \model as it will have more training examples to localize the knowledge obtained during pre-training and transfer learning.

Third, we relied upon the trace matrices provided by each of the four project datasets to serve as training data for TLC and TLC tracing tasks, and as answer sets for evaluating results. However, any inaccuracies in these matrices introduce noise for training purposes and also could introduce inaccuracies in the metric results.

Finally, throughout our study we made strategic decisions about which models to use for baselines, how to build the domain and project corpora, and which adjacent tasks to evaluate in our transfer learning processes. While alternate approaches might unearth different `winners', we observed trends that confirmed our conjecture that adjacent tasks related to matching sentences and were therefore quite similar in nature to the tracing task performed better than other types of tasks.  We leave further investigation of tasks to future work, and release our data to facilitate further studies.

\section{Related work}
\label{tse_sec:related_work}
The classification and sentence embedding models are the most commonly used architectures for automating the creation of trace links. Classical approaches built using information retrieval techniques belong to the Sentence Embedding category. Tracing models such as VSM \cite{salton1975vector}, LDA\cite{blei2003latent} and LSI\cite{papadimitriou2000latent} analyze the common terms that are shared between source and target artifacts to determine the likelihood of a potential link; however, these methods are generally unable to produce accurate links on large scale projects due to the semantic gap between artifacts \cite{guo2017semantically}. To address this problem,  Liu \etal \cite{liu2020traceability} proposed improved VSM models that leverage concept relations from manually created knowledge bases or automatically constructed word embeddings. Researchers also explored machine learning approaches, which mostly fall into the category of classification. For ML methods, feature engineering techniques have been extensively utilized. Heuristic rules have been applied to extract semantic relations between artifacts as semantic features \cite{rath2018traceability}, and then ML models such as Random Forest\cite{riedmiller2014multi} and MaxEnt\cite{berger1996maximum} models have been applied to predict the artifact relevance based on these manually extracted features. Although, these two approaches partially mitigate the semantic gap, the resulting trace model can not actually comprehend the meaning of artifacts. To address this problem, Guo \etal proposed a deep learning tracing model, referred to as TraceNN \cite{guo2017semantically}. Their model applied a bidirectional recurrent neural network \cite{schuster1997bidirectional} to encode the semantic representation of unprocessed NL artifacts. 
It then utilized a multi-layer perceptron (MLP) classification network \cite{riedmiller2014multi} to predict the distance between two encoded artifacts. By taking the surrounding context around the words into consideration they created a semantic representation for a whole artifact. This contrasts with classical IR based methods that do not consider this context.

Since DL approaches based-on LMs have delivered better performance than RNN models on various NLP tasks, Lin \etal proposed LM-based trace models to trace the code change set in commits to issue discussions in open source projects \cite{lin2021traceability}. To allow the LM to understand the grammar of programming languages they built their model using the CodeBert LM \cite{feng2020codebert} which was \pretrained on open source code and documentation. Their results showed that such LM-based models produced more accurate trace links than Guo's TraceNN approach. Our study further investigates the effectiveness of applying LM-based approaches to address the text-to-text tracing task instead of text-to-code. Other LM based trace models, such as DeepMatcher \cite{haering2021automatically}, used DistillBert as an alternative encoder of VSM and RNN; however the output of DistillBert was directly applied to Cosine Similarity without fine-tuning. We argue that this method is relatively weak because it only uses the knowledge from the pretraining stage without the benefit of calibration with the target project through fine-tuning.

Domain specific NLP tasks such as Name Entity Recognition (NER) in a technical corpus \cite{dougan2014ncbi, luan2018multi}, Reference Prediction across academic papers \cite{bird2008acl}, and Construction/Expansion of a domain ontology \cite{kim2003genia, cohan2019structural} also require knowledge about terminology in the document. However, LMs \pretrained on a general corpus does not include many project-specific terms and therefore are not trained on the full vocabulary of the downstream task. To address this problem, researchers have created  domain-specific LMs by collecting a large corpus of data for a target domain, and \pretrained a LM from scratch to include domain terminology. Models such as SciBERT\cite{beltagy2019scibert}, BioBERT\cite{lee2020biobert}, ClinicalBERT\cite{huang2019clinicalbert} and FinBERT\cite{araci2019finbert} have been created to cover the general science/academia, biology, clinical, and finance domain. Tai \etal also proposed an enhanced pretraining framework to reduce the training time for constructing a domain LM \cite{tai2020exbert}.
Another branch of research explores the continual refinement of an existing LM to augment it with domain knowledge. Rongali \etal, \cite{rongali2020continual} and Sun \etal, \cite{ sun2020ernie} have explored methods for augmenting the training of a generic LM with an additional domain corpus. In both approaches, a relatively large sized domain corpus was collected from web scraping or using an additional publicly accessible domain corpus, such as PubMed \cite{canese2013pubmed} or Wikipedia. However, software projects cover a wide range of domains and it is challenging to find, or construct a sufficiently large corpus for each and every project, especially as engineers may introduce  new concepts for specific projects. Our study explores pretraining of a LM by augmenting the domain corpus with  project artifacts.
\section{Conclusion}
\label{tse_sec:conclusion}
In this study, we have proposed \model as a technique for completing, expanding and generating text-to-text trace links. \model leverages a LM as its underlying knowledge base and aims to produce trace links with a high degree of accuracy. Our experimental results show that \model can generally outperform classical IR trace models when a set of training links are available that allow it to take advantage of transfer learning techniques. 
Specifically, our work compared three  transfer learning strategies using different knowledge sources. 
First, we collected data from GitHub and formulated a  corpus that was used to pretrain an LM dedicated to software projects.  Our results showed that this improved the tracing performance in comparison to general LMs, and performed particularly well when the project belonged to a domain with an active GitHub community.
Second, we explored domain adaption of an LM by extending its pretraining using a project dedicated corpus that was retrieved in an automated manner using a data pipeline developed in this work. The corpus was built specifically around the concepts found in the targeted project. Building the corpus required significantly less time and computing resources than training a new LM from scratch, but still improved over the performance of more generic domain LMs, such as SciBert, that had previously been \pretrained using tens of thousands of topical papers. 
Third, we mined the AutoLinks from GitHub and used them to create a closely adjacent tracing task which supported task-level transfer.  This strategy achieved the best performance overall, even for the harder problem of generating links when little or no training data was available (i.e., the TLG tracing task). In our experiments, when given even 10 training examples, it was able to outperform classical IR models. 

While the results achieved in this work are not yet perfect, they represent significant improvements over existing tracing techniques.  To put the results in perspective, based on our previous experience of traceability in practice, we estimate that for full industry adoption, we need to achieve MAP and F2 scores with an accuracy of 0.8 or higher \cite{DBLP:journals/computer/Cleland-HuangBCSR07}. These results get our closer and additional improvements are clearly achievable. Furthermore, the ever increasing size of the training corpus and the subsequent knowledge provided by baseline LMs, such as GPT-3 \cite{brown2020language}, suggests that accuracy will continue to improve at a rapid pace, and that effective automated traceability solutions are within  reach to support traceability in large and diverse industrial applications.

This paper has provided an initial exploration of how deep learning and transfer learning can be used to leverage these baseline LMs and to address the requirements traceability problem. However, there is still much research to be performed, and the results and observations from this study suggest several compelling future research directions. These include the exploration of combining multiple transfer learning strategies where different knowledge sources can harmoniously collaborate to improve the downstream tracing task through multi-task learning. 

Another direction focuses on the model architecture. The CSE architecture has already shown great potential for software traceability though in our experiments it did not  outperform the CLS architecture.  However, in comparison to CLS, the CSE architecture is more scalable for larger projects because it does not require a classification network, which creates an efficiency bottleneck for CLS by its need to compute a similarity score using a formula such as the simple Cosine Similarity. However, the existing SimCSE framework is designed for sentence semantic alignment in general NLP and has not been optimized for software traceability. By improving the self-supervision heuristic and negative sampling strategies, we hypothesize that such a framework could potentially reduce the noise that is currently inherent to training examples and ultimately outperform CLS approaches.

\vspace{5pt}
\noindent\textbf{Open Science:} All models and source code, as well as the generated monitoring infrastructures, are available for review purposes (but not yet advertised for public use) at the following google drive folder: \url{https://drive.google.com/drive/folders/1EPAtwWI8BBZVi-NQj7W5J549SnD69HHy?usp=sharing}
 We will move them to a permanently archived site upon acceptance of this paper.

\ifCLASSOPTIONcompsoc
  \section*{Acknowledgments}
\else
  \section*{Acknowledgment}
\fi

The work described in this paper was partially funded by the USA National Science Foundation under grant $\#$CCF-1901059.

\balance
\ifCLASSOPTIONcaptionsoff
  \newpage
\fi

\bibliography{bibs/lm_trace,bibs/tse2022,bibs/pfi}
\bibliographystyle{IEEEtran}




%








\end{document}